\documentclass{sig-alternate-05-2015}

\CopyrightYear{2017}
\setcopyright{acmcopyright}
\conferenceinfo{WSDM 2017,}{February 06-10, 2017, Cambridge, United
Kingdom}
\isbn{978-1-4503-4675-7/17/02}\acmPrice{\$15.00}
\doi{http://dx.doi.org/10.1145/3018661.3018672}

\usepackage{color}
\usepackage{verbatim} 
\usepackage{subfigure} 
\usepackage{multirow}
\usepackage{algorithm}
\usepackage[noend]{algorithmic}
\usepackage{xspace}
\usepackage{graphicx}
\usepackage{epsfig}
\usepackage{amsmath}
\usepackage{amssymb}
\usepackage{times}
\usepackage{float}
\usepackage{enumitem}
\usepackage{balance}
\usepackage[font=bf,belowskip=0pt,aboveskip=5pt]{caption}

\usepackage{booktabs} 

\newcommand{\hide}[1]{}
\newcommand{\xhdr}[1]{\vspace{1.7mm}\noindent{{\bf #1.}}}
\newcommand{\xhdrnodot}[1]{\vspace{1.7mm}\noindent{{\bf #1}}}

\newcommand{\etc}{\emph{etc.}}
\newcommand{\eg}{\emph{e.g.}}
\newcommand{\ie}{\emph{i.e.}}

\usepackage{etoolbox}
\newcommand{\zerodisplayskips}{%
  \setlength{\abovedisplayskip}{4pt} 
  \setlength{\belowdisplayskip}{4pt}
  \setlength{\abovedisplayshortskip}{4pt}
  \setlength{\belowdisplayshortskip}{4pt}}
\appto{\normalsize}{\zerodisplayskips}
\appto{\small}{\zerodisplayskips}
\appto{\footnotesize}{\zerodisplayskips}

\let\oldenumerate\enumerate
\renewcommand{\enumerate}{
  \vspace{-0.7\topsep} 
  \oldenumerate
  \setlength{\itemsep}{1pt}
  \setlength{\parskip}{0pt}
  \setlength{\parsep}{0pt}
  \setlength{\topsep}{0pt}
  \setlength{\partopsep}{0pt}
}
\let\olditemize\itemize
\renewcommand{\itemize}{
  \vspace{-0.7\topsep}
  \olditemize
  \setlength{\itemsep}{1pt}
  \setlength{\parskip}{0pt}
  \setlength{\parsep}{0pt}
  \setlength{\topsep}{0pt}
  \setlength{\partopsep}{0pt}
}

\graphicspath{{./FIG/}}

\begin{document}

\title{Online Actions with Offline Impact: How Online Social Networks Influence Online and Offline User Behavior
}

\numberofauthors{3}
\author{
\alignauthor Tim Althoff\\
      \affaddr{Stanford University}\\
      \email{althoff@cs.stanford.edu}
\alignauthor Pranav Jindal\\
      \affaddr{Stanford University}\\
      \email{pranavj@cs.stanford.edu}
\alignauthor Jure Leskovec\\
      \affaddr{Stanford University}\\
      \email{jure@cs.stanford.edu}
}

\maketitle

\begin{abstract}


Many of today's most widely used computing applications utilize social networking features and allow users to connect, follow each other, share content, and comment on others' posts. 
However, despite the widespread adoption of these features, there is little understanding of the consequences that social networking has on user retention, engagement, and online as well as offline behavior. 

Here, we study how social networks influence user behavior in a physical activity tracking application. 
We analyze 791 million online and offline actions of 6 million users over the course of 5 years, and show that social networking leads to a significant increase in users' online as well as offline activities. 
Specifically, we establish a causal effect of how social networks influence user behavior. 
We show that the creation of new social connections increases user online in-application activity by 30\%, user retention by 17\%, and user offline real-world physical activity by 7\% (about 400 steps per day).
By exploiting a natural experiment we distinguish the effect of social influence of new social connections from the simultaneous increase in user's motivation to use the app and take more steps. 
We show that social influence accounts for 55\% of the observed changes in user behavior, while the remaining 45\% can be explained by the user's increased motivation to use the app.
Further, we show that subsequent, individual edge formations in the social network lead to significant increases in daily steps.
These effects diminish with each additional edge and vary based on edge attributes and user demographics.
Finally, we utilize these insights to develop a model that accurately predicts which users will be most influenced by the creation of new social network connections.

\end{abstract}




\section{Introduction}
\label{sec:intro}

Social network features are central to many of today's computing applications. 
Many successful websites and apps use social networking features to appeal to their users, allowing them to interact, form social connections, post updates, spread content, and comment on other's posts. Social networking features are ubiquitous and are not only used by online social networks, such as Facebook and Twitter. For example, news reading, online education, music listening, book reading, diet and weight loss, physical activity tracking, and many other types of modern computing applications all heavily rely on social networking.

Recent research 
has made great advancements towards understanding of
fundamental structural properties~\cite{leskovec2008planetary,mislove2007measurement},
growth~\cite{leskovec2008microscopic},
navigability~\cite{kleinberg2001small,leskovec2014geospatial},
community structure~\cite{backstrom2006group,yang2012defining},
information diffusion~\cite{bakshy2012role,cheng2014can},
influence maximization~\cite{kempe2003maximizing},
social capital~\cite{burke2010social,ellison2007benefits},
and social influence~\cite{Salganik2006} in online social networks.
However, the impact of the online social networks on user behavior remains elusive.
For example, little is known about whether and to what degree online social networking features influence user engagement, increase user retention, and change behavior within the immediate application as well as in the real-world. 
Furthermore, it is not clear whether social networking features simply attract users that would be more active and more engaged even if these features were absent, and whether social networks actually influence user online as well as offline behavior.

Existing studies of social influence in social networks have mainly been restricted to measuring online behaviors and outcomes such as the adoption of apps~\cite{aral2009distinguishing,Aral2012susceptible}, 
downloads of content~\cite{Salganik2006}, voting on content~\cite{muchnik2013social}, and resharing of content~\cite{myers2012information,romero2011differences}. 
However, many important behaviors and outcomes pertain to the offline world including political mobilization~\cite{Bond2012facebook}, physical activity~\cite{althoff2016quantifying,althoff2016pokemon,dishman1996increasing,Lee2012pa}, food intake~\cite{de2016characterizing}, mental health~\cite{althoff2016counseling,de2016discovering}, obesity~\cite{Christakis2007obesity}, and smoking~\cite{Christakis2008smoking}.
In order to study offline behaviors, researchers have used proxies that are observable online: for example, posting in a particular forum or an app to measure dieting choices~\cite{de2016characterizing,mejova2015foodporn}, suicidal thoughts~\cite{de2016discovering}, helping behavior~\cite{althoff2014howtoaskforafavor} and charitable behavior~\cite{althoff2015donor}. 
However, one has to trust that the self-reports observed online 
correspond to objective behaviors and many studies have shown large biases of such self-reports~\cite{belli1999reducing,hebert1995social,Tucker2011selfreport}.

Estimating the influence of social networks on online as well as offline behavior is challenging due to unobserved counterfactual behavior, where one cannot observe a user's behavior had they not joined the social network.
Furthermore, selection effects complicate causal estimation from observational data~\cite{anagnostopoulos2008influence,aral2009distinguishing,la2010randomization}. For example, social network users could exhibit different behaviors due to (1) a selection effect of what kind of users would choose to join the social network, or (2) an actual influence effect of the social network on their behavior. 
Often the mere act of being part of a social network 
already means that these users are particularly motivated to use the app and take more steps. In many contexts all of these effects are acting simultaneously (\eg, \cite{Bauman2002} for health behaviors), which creates further challenges and makes causal identification of effects even harder.
 
\xhdr{Present work}
In this paper, we study the influence of social networks on users' online as well as offline behavior. 
We study user behavior in a smartphone physical activity tracking application, that allows us to observe users' in-application engagement as well as the offline real-world physical activity measured through smartphone accelerometry. 
Specifically, we use data from the Azumio Argus app, which tracks exercise and physical activity of 6 million users worldwide over the course of 5 years (2011-2016). 
During this time, users created 631 million activity posts (\eg, runs, sleep, cycling, yoga, \etc) by actively opening the app and self-reporting the activity.
In contrast, physical activity is passively collected through smartphone sensors in the form of step measurements without the need for self-reports and active user engagement.
Our data set additionally contains 160 million days of passive steps tracking adding up to 824 billion total steps taken.
Therefore, we distinguish between activity posts as a measure of online app engagement and steps taken as a measure of offline physical activity behavior.
To the best of our knowledge this is the largest dataset on human activity tracking and social network interactions to date (\eg, ten thousand times more users and a million times more activity tracking than comparable studies \cite{ebrahimi2016characterizing}).

An internal social network was introduced in the application in November 2013 and since then a subset of users have chosen to join and engage in the social network. 
Using the data, we quantify the causal effect of the social network on user behavior by harnessing a novel natural experiment on delayed social network edge formation. 
In particular, we distinguish the causal effect of social influence of a new network connection from the simultaneous increase in motivation of the user to use the app (\ie, a selection effect). 
We show that social influence does explain 55\% of the observed average effect, while the remaining 45\% of the observed effects are due to the increased motivation.

We find that joining the social network has significant positive effect on online and offline user behavior that diminish over time. 
Users of the social network are 30\% more active in the app, 17\% less likely to drop out of the app within one year, and 7\% more physically active compared to a matched control group, and we show that these effects last over long periods of several months.
%
Further, we estimate the effect of subsequent, individual edge formations in the social network.
We observe temporary increases in offline physical activity that diminish with each additional connection and 
are larger for friend connections than follower connections.
Further, these average increases 
are larger for the initiator of the connection than its recipient,
and the effect varies with age, gender, weight, and prior physical activity level.
Finally, we utilize these insights to develop a model to predict which users will be most influenced by the creation of new social network connections and show that the proposed factors explain a significant fraction of the variability in user's behavior change.
We conclude by discussing related work on online social networks and social media in the context of health applications.

In summary, the main contributions of this work include:
\begin{enumerate}
\item We study the causal impact of social network features on user behavior using the largest activity tracking dataset to date.
\item We show how online social networks shape online as well as offline user behavior, including user engagement, retention, and real-world physical activity.
\item We employ natural experiments, difference-in-difference models, and matching-based observational studies to disentangle selection effects from causal social network effects.
\end{enumerate}
While we focus on physical activity and health behaviors in this work, our methods are more generally applicable to other offline and online activities.

\pagebreak
\section{Dataset Description}
\label{sec:dataset}

We use a dataset of 6 million individuals from over 100
different countries using the Argus smartphone app by Azumio which allows users to track their daily activities.
Over a time period of 5 years we observe 631 million self-reported activity posts (including running, walking, sleep, heart rate, yoga, cycling, weight, \etc) 
and 160 million days of steps tracking (objectively measured through the smartphone accelerometers) between January 2011 and January 2016. 
Table~\ref{tab:dataset_statistics} further summarizes the worldwide dataset and shows that the distribution of age, gender, and weight is fairly representative of the overall population in many developed countries. 
For example, the median age in our dataset for U.S. users is 34 years which is close to the official estimate of 37 years.
Furthermore, 28\% of these users are obese closely matching previous published estimates of 30-38\%.
We also include plots of the degree distribution and distribution of edge inter-creation times in the online appendix~\cite{althoff2016onlineappendix}.

Throughout the paper, we distinguish between the number of accelerometer-defined steps (physical activity; offline) and the number of posts the user creates within the app each corresponding to a self-reported action such as running, cycling or sleeping (in-app activity; online).
We use ``activity'' to refer to offline physical activity and ``posts'' to refer to online in-application user activity.

In November of 2013, a social network feature was introduced in the app allowing both bi-directional friend connections (after approval of a friend request by the receiver) as well as uni-directional follower connections (without need for approval). 
New social connections result in receiving notifications of the other person's activity posts (\eg, runs and walks).
Furthermore, the app includes a timeline-like activity feed that then includes the activity of the new friend and enables the user to comment on others' activity posts for encouragement and support.
During our observation period, all edges in the network were created organically without any influence of friend recommendation algorithms.

This dataset uniquely enables the study of how social network features impact user behavior. 
It captures both online (actions within the app) as well as offline user behaviors (physical steps taken in the offline world).
Further, it has two relevant properties:  
(1) The social network was introduced after two years of observing behavior without any social interactions. 
(2) The delays with which friendship requests get accepted form a natural experiment, which allows us to disentangle influence of social networking features from simple selection effects.
Our analyses exploit these properties to carry out two natural experiments that provide novel insights into how user behavior is shaped by social network interactions.
Lastly, the large-scale nature of our dataset---about two million times more posts than previously published research~\cite{ebrahimi2016characterizing}---allows us to study various kinds of heterogeneous effects, for example across age, gender, BMI, previous activity level.
Data handling and analysis was conducted in accordance with the guidelines of the appropriate Institutional Review Board. 

\begin{table}[t]
\centering
 \small 
\resizebox{.99\columnwidth}{!}{%
\begin{tabular}{ll}
  \toprule
  Observation period & Jan. 2011 -- Jan. 2016\\
  Introduction of the social network & November 2013\\
  \# total users & 6.0 million \\
  \# total online and offline activities & 791 million\\
  \# activity posts (online engagement) & 631 million\\ 
  \# users tracking steps & 2.0 million\\
  \# days of steps tracking (offline activity) & 160 million\\
  \# total steps tracked & 824 billion\\
  \# users in the social network & 211,383\\
  \# edges in the social network & 563,007\\
  Median age & 33 years\\
  \% users female & 46.1\%\\
  \% underweight (BMI < 18.5) & 4.7\%\\
  \% normal weight (18.5 $\leq$ BMI < 25) & 44.2\%\\
  \% overweight (25 $\leq$ BMI < 30) & 30.2\%\\
  \% obese (30 $\leq$ BMI) & 20.9\%\\
  \bottomrule
 \end{tabular}
 }
 \caption{
 Dataset statistics. 
 BMI refers to body mass index.
 }
 \label{tab:dataset_statistics}
 \end{table}

\section{Distinguishing Intrinsic Motivation from Social Influence}
\label{sec:intrinsic_motivation}

In this section, we quantify the average influence that a single social network edge has on the person who sends the friendship request.
We present a novel approach based on a natural experiment to distinguish the effect of 
(1) increased user's motivation when adding a new edge, from
(2) establishing a social network edge that influences the user to change their behavior.

\begin{figure}[t]
\centering
  \includegraphics[width=.85\columnwidth]{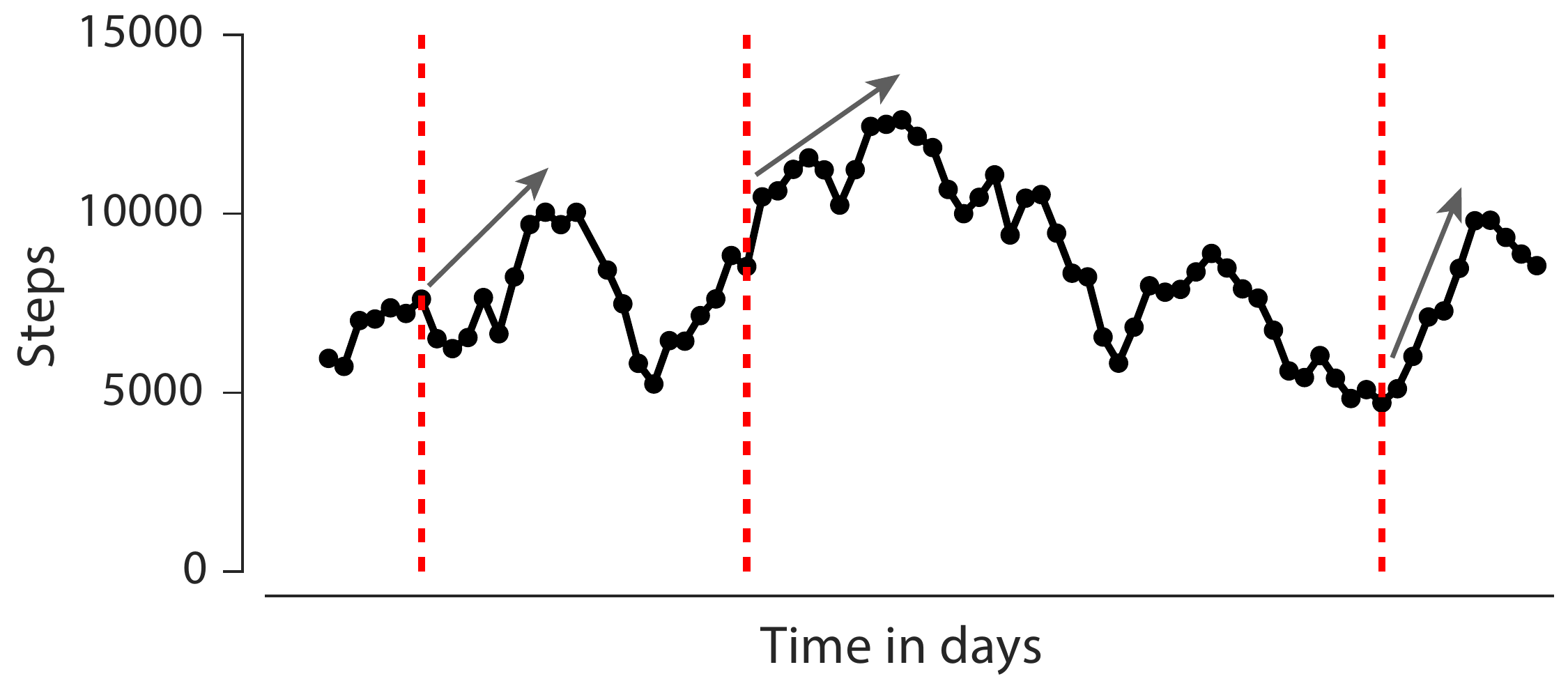}
  \vspace{-2mm}
  \caption{
  Time series of daily steps for an example user. 
  Dashed vertical lines correspond to edge creations.
  We observe significant increases in activity after each created edge (arrows).
  }
  \label{fig:example_timeseries}
\end{figure}

\xhdr{Challenges of distinguishing motivation from influence}
To illustrate the challenge of distinguishing the two effects, Figure~\ref{fig:example_timeseries} shows the daily activity of a user measured by the number of steps over time. Red dotted lines represent the times when the user created new social network connections. Notice that every new edge seems to come with significant increases in activity (arrows) and one could (perhaps mistakenly) see the increase in physical activity after the edge creation as a sign of social influence of the friendship. 

However, it is also possible that there is a selection effect; that is, 
users adding a new edge could simply be more motivated to be active.
They might be particularly excited about the app and activity tracking and inclined to use it even more in the future.
For example, notice that the user in Figure~\ref{fig:example_timeseries} is already increasing her activity before the addition of the second edge which could indicate that she was already motivated to further increase her activity independent of the edge creation.

Such intrinsic motivation could often occur at the same time as the creation of an edge since users might also use the app and its social features more when they are particularly motivated. Thus, one might mistakenly attribute any behavior change to the edge creation instead of the change in user's intrinsic motivation.

\xhdr{Our approach: A natural experiment}
The key idea of our approach is to identify a natural experiment where 
we control for the amount of intrinsic motivation and vary the effect of social edge formation.
We achieve this by relying on the fact that social network edges only become active (\ie, expose users to notifications and posts), after being accepted by the target person. 
This means we can compare the population of users who all sent edge requests (\ie, they were all intrinsically motivated) but some of these requests get accepted immediately, while others get accepted with significant delay. 
By comparing the activity levels of senders of edges which get accepted immediately versus late, we can distinguish what fraction of activity is due to social network influence and what fraction is due intrinsic motivation\footnote{Note that we do not consider the receiver of an edge request particularly motivated at the time of the request since receiving an edge request does not require any action or increased motivation. Thus, we focus our study on users sending requests only.}.
Conceptually, delayed acceptance of edges forms a natural experiment~\cite{rosenzweig2000natural} where the treatment of establishing a new friendship is assigned ``as if random'' among users who chose to add a new a friend.
To the best of our knowledge this is the first time delayed edge formation is used to identify social influence effects.

Next, we describe how this natural experiment allows us to distinguish the effect of intrinsic motivation from social influence of a single social network edge.
We then proceed to show that delayed acceptance of friendships seems indeed random.

\begin{figure}[t]
\centering
\includegraphics[width=.85\columnwidth]{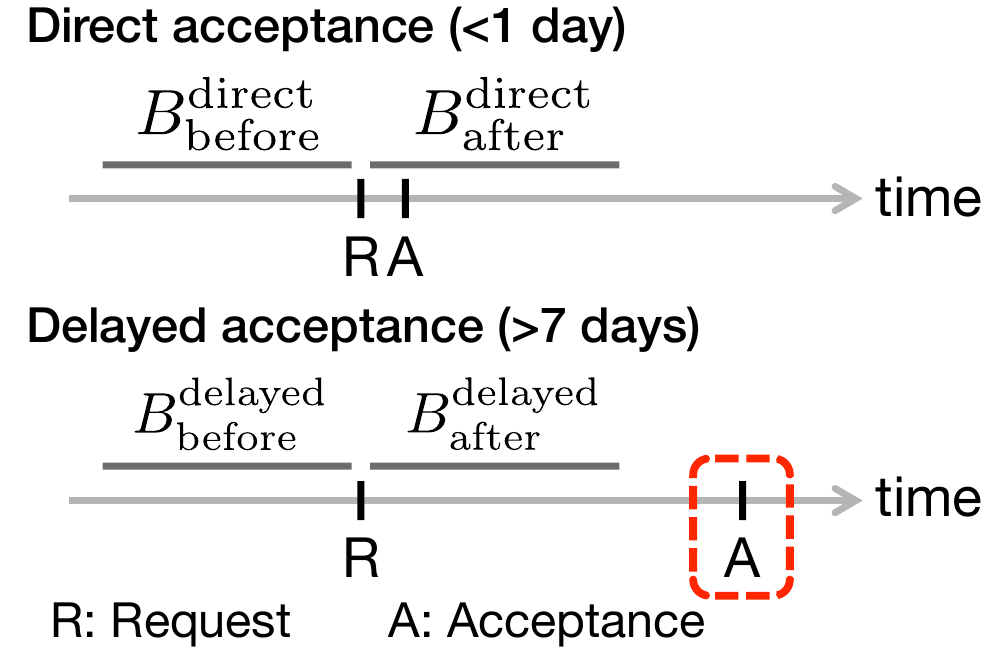}
\caption{
Conceptual framework using delayed accepted edges to disentangle social influence due to edge creation and intrinsic motivation.
$B_{\text{before}}$ and $B_{\text{after}}$ to refer to the user behavior 7 days before and 7 days after the friendship request (R) for the user who sent the friend request.
Delayed accepted edges get accepted (A) after this period (more than 7 days after the request).
Directly accepted edges are accepted (A) within one day of the request (R).
}
\label{fig:delayed_accepted_concept_figure}
\end{figure}

\xhdr{Conceptual framework}
We define \textit{delayed acceptance} as any request that gets accepted after more than 7 days by the recipient of the request.
We compare the difference in physical activity levels 7 days after compared to 7 days before the time of the friendship request for both directly accepted (within 1 day of the request; 81\% of all friend requests) 
and delayed accepted friend requests (at least 7 days after the request; 8\% of all friend requests; median time to acceptance of 25 days).
We use $B_{\text{before}}$ and $B_{\text{after}}$ to refer to the user behavior before and after the friendship request for the user who sent the friend request.
The conceptual framework for our analysis is illustrated in Figure~\ref{fig:delayed_accepted_concept_figure}.
While this framework generally applies to any user behavior, we use accelerometer-recorded daily steps here because other activities (\eg, yoga or weightlifting) are self-reported and potentially biased.

This framework allows us to disentangle intrinsic motivation $M$ from social influence $I$.
First, we consider the difference in behavior within an individual around the time of the friendship request. 
For directly accepted edges, the difference in behavior could come from both increased motivation $M$ but also from social influence $I$ since the connection indeed is made and the two friends can now observe and get notified on each other's activity.
Mathematically, this corresponds to $B_{\text{after}}^{\text{direct}} - B_{\text{before}}^{\text{direct}} = M + I$.
For delayed accepted edges, the difference in behavior can only stem from the increased motivation $M$ but not from social influence $I$ since the connection does not get created until at least 7 days later (\ie, the edge creation happens after the $B_{\text{after}}^{\text{delayed}}$ period).
Therefore, we have $B_{\text{after}}^{\text{delayed}} - B_{\text{before}}^{\text{delayed}} = M$.
We can identify the social influence effect $I$ by taking the difference between those differences:
$(B_{\text{after}}^{\text{direct}} - B_{\text{before}}^{\text{direct}}) - (B_{\text{after}}^{\text{delayed}} - B_{\text{before}}^{\text{delayed}} ) = (M + I) - (M) = I$.
This corresponds to a difference-in-difference analysis well known in the econometrics literature~\cite{Lechner2011diffindiff}.
We estimate these factors as the average over all accepted edges (N=34,324) and delayed accepted edges (N=3,146) excluding any edges that do have other edges within a 7 day window around the time of request.
This means that during $B_{\text{before}}^{\text{direct}}$, $B_{\text{after}}^{\text{direct}}$, $B_{\text{before}}^{\text{delayed}}$, and $B_{\text{after}}^{\text{delayed}}$, there are no other edges being created that could bias the effect estimate.
Note that there are no issues of serial correlation which can lead to underestimating the variance of difference-in-difference estimates when many time steps are used~\cite{bertrand2004much} (\S 4C). 

\begin{table}[t]
\centering
\small 
\resizebox{.90\columnwidth}{!}{%
\begin{tabular}{llr} 
  \toprule
    \textbf{Group} &  \textbf{Variable} & \textbf{SMD} \\ 
   \midrule
  \textbf{Sender} & Age & 0.092 \\ 
      & Age NA & 0.071 \\ 
      & Gender & 0.115 \\ 
      & Gender NA & 0.105 \\ 
      & BMI & 0.052 \\ 
      & BMI NA & -0.013 \\ 
      & Steps 7 days before & 0.034 \\ 
      & Days tracked 7 days before & -0.005 \\ 
  \textbf{Receiver}  & Age & -0.004 \\ 
      & Age NA & -0.061 \\ 
      & Gender & -0.026 \\ 
      & Gender NA & -0.028 \\ 
      & BMI & -0.034 \\ 
      & BMI NA & -0.065 \\ 
      & Steps 7 days before & 0.074 \\ 
      & Steps 7 days before NA & -0.002 \\ 
      & Days tracked 7 days before & -0.014 \\ 
  \textbf{Relationship} & \#Mutual friends at request & 0.080 \\ 
  \textbf{Timing} & Edge number for sender       & 0.049 \\ 
      & Edge number for receiver       & 0.098 \\ 
      & \#Days on social network for sender & 0.109 \\ 
      & \#Days on social network for receiver & 0.179 \\ 
  \midrule
   & \textbf{Median Absolute SMD} & 0.057 \\ 
   & \textbf{Maximum Absolute SMD} & 0.179 \\ 
   \bottomrule
 \end{tabular}
 }
 \caption{
 Balancing statistics on relevant covariates for the natural experiment.
 Covariates with absolute SMD lower than 0.25 are considered balanced (\eg,~\cite{stuart2010matching}).
 NA refers to missingness indicator.
 \#Days on social network refers to the number of days between the first created edge and the friendship request.
 }
 \label{tab:natural_experiment_balancing}
 \end{table}

\begin{figure}[t]
\centering
\includegraphics[width=.80\columnwidth]{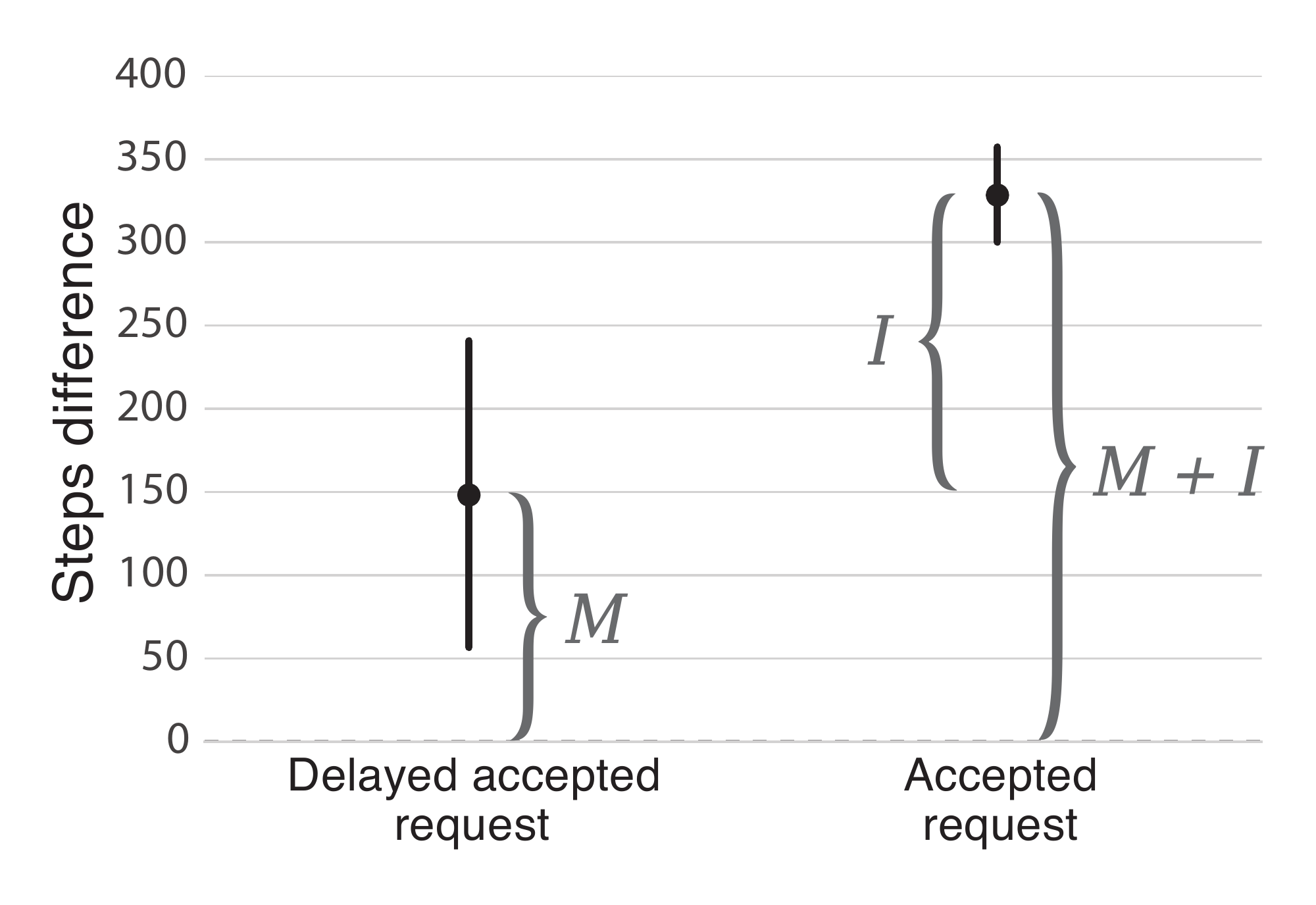}
\vspace{-3mm}
\caption{
Steps difference after the time of friendship request for delayed accepted and directly accepted friendship requests.
Delayed accepted requests lead to significant increases in activity (148 additional daily steps), even before they get accepted.
However, intrinsic motivation $\mathbf{M}$ only explains 45\% of the observed effect for directly accepted requests.
We attribute the remaining 55\% or 180 daily steps to social influence $\mathbf{I}$.
Error bars in this and all following plots represent bootstrapped 95\% confidence intervals.
}
\label{fig:delayedaccepted_vs_accepted_steps}
\end{figure}

\xhdr{Validating randomization assumption}
Next, we establish that the mechanism underlying our natural experiment (that is, the assignment of friend requests to delayed acceptance versus direct acceptance) seems indeed random. For example, it could be the case that late acceptance of a friend request signifies a weak friendship, while fast acceptance might indicate a strong friendship which could also have a stronger social influence.
However, if the assignment of delayed vs. direct acceptance is random, we would expect both groups to be indistinguishable based on any properties of the requests.
Two groups are called indistinguishable or \textit{balanced}, if all covariates are within a standardized mean difference (SMD) of 0.25 standard deviations (\eg,~\cite{stuart2010matching})
\footnote{
Note that SMD is preferred over hypothesis tests and p-values as a measure of balance since the latter conflate changes in balance with changes in statistical power~(\eg,~\cite{stuart2010matching}).}. 
Standardized mean difference is defined as the difference in means of treated and control groups divided by the standard deviation within the treated group~\cite{stuart2010matching}.
%
In the following, we consider a large number of relevant covariates that, if any were imbalanced between delayed accepted and accepted requests, would cast doubt on the randomness of our natural experiment mechanism.
We further include balance checks on binary variables indicating missingness (NA) in any of the covariates (recommended in~\cite{rosenbaum2010design} \S 9.4).
However, we will show that our mechanism creates two groups that are indeed well balanced across all these variables and can be considered random. 

We consider the balancing of the following covariates summarized in Table~\ref{tab:natural_experiment_balancing}.
First, the \textit{users sending} the delayed accepted friend request could be fundamentally different from users sending a friend request that gets accepted directly.
Different users might behave differently around the edge creation and we need to make sure that any measured differences between the two groups are due to the edge creation and not due to other factors.
Second, the same could be true for the \textit{users receiving} the friend request.
However, we find both the users sending as well as the users receiving the friend request to be very well balanced (all SMD significantly smaller than 0.25) on age, gender, BMI, and pre-treatment behavior using average number of steps 7 days before the friend request and the number of days tracked during that window.
Third, there could be a difference in the \textit{relationship} between the sending and receiving users of delayed accepted and accepted edges.
For example, users accepting late could be weak ties, while fast accepting users might be strong ties.
Still, we find the number of mutual friends between the two users at the time of the request to be well-balanced (SMD=0.080) 
indicating that this is not the case.
Lastly, the \textit{timing} of accepted and delayed accepted requests within the user lifetime could be a distinguishing factor, and senior users might behave differently from junior users of the app.
Nevertheless, we find the node degree at request time and the time elapsed since joining the social network for both sender and receiver to be balanced as well (all SMD < 0.25).
Overall, all covariates are balanced within 0.179 standard deviations and half of the covariates are within 0.057 standard deviations which is much lower than the typical threshold of 0.25 standard deviations.
Since all observed covariates are well balanced we consider the natural experiment mechanism to be as if random and that therefore unobserved covariates would also be reasonably well balanced and not confounding our effect estimate.

\xhdr{Results}
The difference in post-treatment activity for both groups is shown in Figure~\ref{fig:delayedaccepted_vs_accepted_steps} (\ie, $B_{\text{after}}^{\text{delayed}} - B_{\text{before}}^{\text{delayed}}$ and $B_{\text{after}}^{\text{direct}} - B_{\text{before}}^{\text{direct}}$).
We find that users increase their average activity by 148 daily steps even after delayed accepted friend requests that are accepted only after our steps observation period ($M = 148$ daily steps difference; significantly larger than 0 according to Wilcoxon signed rank test; $p<10^{-3}$). 
This suggests that users sending a friend request are indeed particularly motivated at the same time.
However, we find that the effect for accepted friend requests is 122\% larger at 328 daily steps ($M+I=328$ daily steps difference compared to pre-treatment; significantly larger than 0 according to Wilcoxon signed rank test; $p<10^{-15}$). 
This means that the difference of 180 average daily steps can be attributed to an influence effect $I$ from the social connection.
This influence effect $I$ is statistically significant (95\% confidence interval between 74 and 236 steps; $p<10^{-3}$; Mann--Whitney U test). 

Becoming friends within the app results in exposure to the friend's activity and status updates through notifications and the personalized activity feed.
The result shows that social influence through these mechanisms indeed leads to higher physical activity levels during the week after establishing the friendship.
To further verify the result we have additionally conducted a matching study between accepted edges and \textit{pending} edges which never got accepted finding very similar effect sizes (see online appendix~\cite{althoff2016onlineappendix}). 

Overall, we estimate that 55\% (180/328) of total behavior change is due to social influence and that 45\% (148/328) is due to the user's elevated intrinsic motivation. 
This experiment establishes a causal effect of a social network connection on offline user behavior.

\section{How Joining a Social Network Impacts User Behavior}
\label{sec:joining_network}

In Section~\ref{sec:intrinsic_motivation}, we quantified the social influence effect of an \textit{average} edge.
In this section, we study the effect of the \textit{very first} edge of a user; that is, the effect of \textit{joining} the network.
Since the social network was introduced only after two years into our observation period, we can measure the effect its introduction had on user behavior.
We show that joining a social network makes users more physically active for a period of several months, 
increases their number of posts within the app, 
and makes them more likely to continue to use the app 
compared to users in a matched control group.

\subsection{User Physical Activity Level}
\label{subsec:joining_steps}
First, we consider whether joining the social network makes users physically more active and estimate the size of this effect over time.

\xhdr{Method}
We compare users that join the social network (treatment) to similar users that do not join the social network (control) both before and after the matched treatment user joins the social network (\ie, difference-in-differences design~\cite{Lechner2011diffindiff} with a matched control group~\cite{rosenbaum2010design,stuart2010matching}).
We define joining the social network as the point in time when a user creates the first connection within the network either by sending or receiving an accepted friend edge (bi-directional) or follow edge (uni-directional). 
We will discuss differences between edges of different direction and type in Section~\ref{sec:edge_effects}. 

In this observational study, we only consider users that have recorded activity before and after the analysis period ranging from 4 weeks before joining the social network until 20 weeks after joining.
This ensures that we can compare post-treatment to pre-treatment activity, and that users tracked their behavior from the beginning and do not drop out during the analysis period.
Further, we only consider users that record steps for at least one day in every 7-day window during the analysis period. 
This ensures that for each week, we estimate the effect from the exact same set of users, rather than confounding the estimate by which users record any activity in a given week.
Since treatment users are active on the day of joining the social network, we further constrain all users 
to have recorded steps on the day of joining to create a fair comparison.

We match each user receiving treatment to a similar control user that did not receive treatment based on the following covariates in an almost-exact matching strategy~\cite{rosenbaum2010design}.
We match on the day of sign-up on the app to control for user lifetime ($\pm$ 14 days).
Further, we match on pre-treatment behavior; that is, the weekly average number of daily steps during each of the four weeks before treatment ($\pm$ 600 steps).
This is important since treatment users might be particularly active overall and also could become more active before joining the network 
since we observed intrinsic motivation effects in Section~\ref{sec:intrinsic_motivation}.

\begin{figure}[t]
\centering
\includegraphics[width=.99\columnwidth]{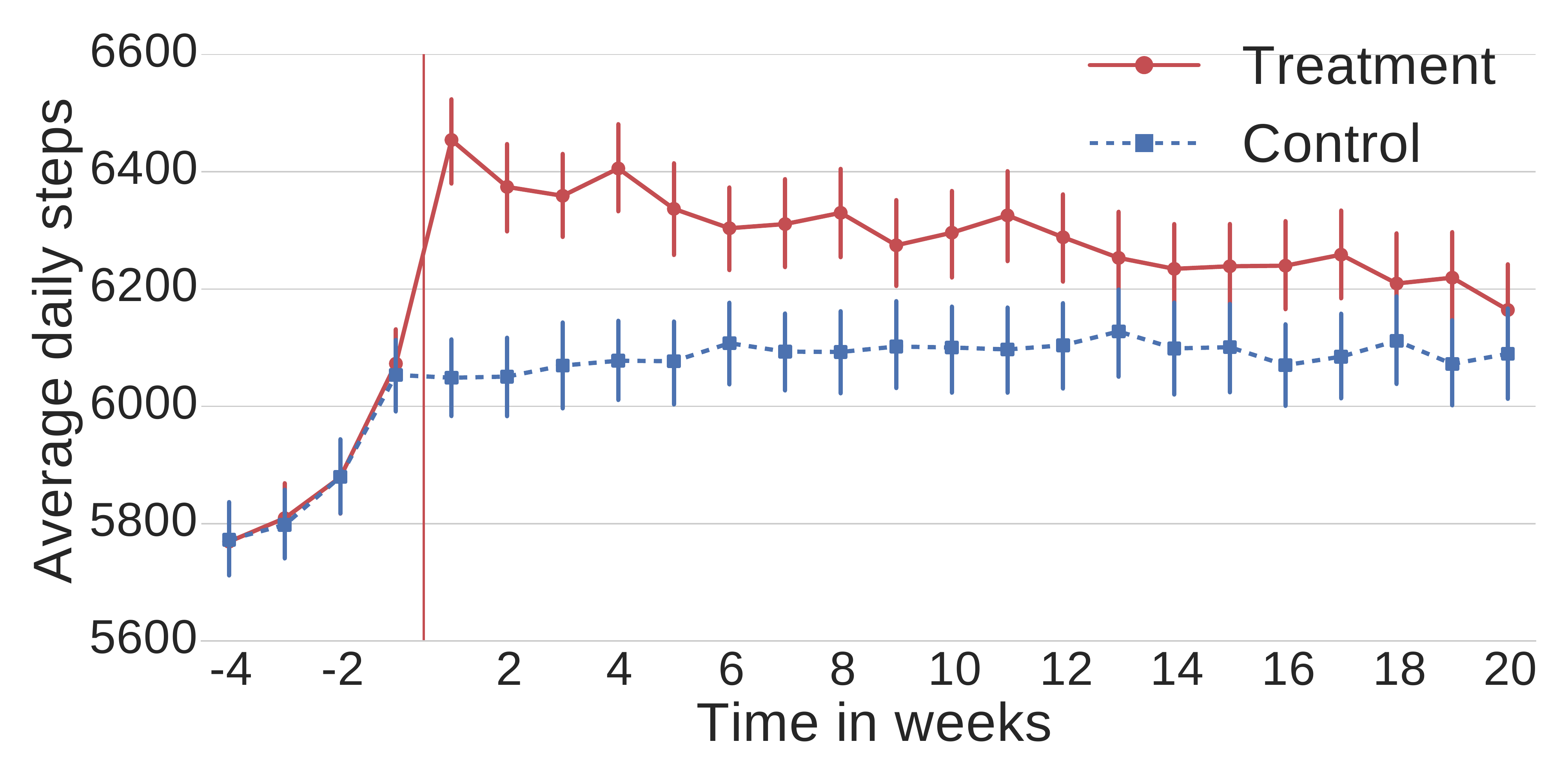}
\vspace{-2mm}
\caption{Average daily steps for users that do join the social network at time zero (treatment; red) and matched users that do not (control; blue). 
We observe a significant boost in activity of 406 additional daily steps in treatment users that diminishes over 20 weeks but no difference in control users.
}
\label{fig:joining_network_design4}
\end{figure}

\xhdr{Results}
Matching results in 6076 matched pairs of users fulfilling the constraints explained above.
Figure~\ref{fig:joining_network_design4} shows that treatment (red) and control users (blue) have indistinguishable levels of pre-treatment activity. 
We observe activity increases in treatment users just before they join the social network (week $-1$).
This shows that these users are particularly motivated even before they join the social network consistent with our findings in Section~\ref{sec:intrinsic_motivation}.
Note, however, that the selected control users are increasing their activity by the exact same amount and therefore can be considered similarly motivated.
When treatment users join the social network (vertical red line at week 0), they exhibit a large spike of 406 additional daily steps (7\% increase) compared to control users (6454 vs 6048 mean steps; $p<10^{-15}$ according to two-tailed, Mann--Whitney U test) 
while activity in control users stays the same. 
Notice also the long-lasting effect of joining the social network.
After creating their first social network connection, treatment users are significantly more active for a period of over three months than the matched control group with identical pre-treatment activity patterns.
In contrast, the users in the control group do not change their activity levels during the post-treatment observation period.
The effect diminishes over a period of 20 weeks at which point the group activity levels are statistically indistinguishable again (6164 vs 6089 mean steps; $p=0.095$ according to two-tailed, Mann--Whitney U test).
This shows that social network features have a significant effect on user behavior that can last over several months.
Overall, the effect of adding the first edge is about 400 additional daily steps in the following week, which is consistent with Section~\ref{sec:intrinsic_motivation} (328 additional daily steps; note that here we excluded compounding effects of multiple edges in a short amount of time).

While due to the careful constraints this experiment considers only a subset of 6,076 out of the 211,383 social network users,
these users are very similar to the average social network user (only 563 more daily steps and 5 years older on average).
Repeating the experiment with less stringent constraints (\eg, reducing the post-treatment observation period to four weeks), 
we observe practically identical effect sizes (382 daily steps increase; see online appendix~\cite{althoff2016onlineappendix}). 
This suggests that the reported effects appear to describe fairly general dynamics in the dataset rather than being an artifact of a specific subset of users.

\subsection{Online User Engagement and Retention}
\label{subsec:joining_retention}
We now shift the focus from social network effects on offline physical activity to effects on user retention and app usage; 
that is, whether or not users continue to use the app and how often they use it to post any tracked activities online.

\xhdr{Method}
As in Section~\ref{subsec:joining_steps}, we compare users that did join the social network (treatment) to comparable users that did not join the social network (control) but started using the app around the same time ($\pm$ 7 days).
Matching is done on the weekly average number of posts per day ($\pm$ 0.3 posts) for each of the four weeks before treatment.
To ensure that we only consider currently active users, we further require all users to have at least one post in each of the four weeks before the treatment user joined the social network.
We further only consider users that have a post on the day when the treatment user joins the social network to create a fair comparison.
First, we observe whether a user returns in any given week over the course of one year by creating any post through the app. 
Second, for the users still using the app (\ie, removing users who did not create any posts during a given week), we measure how much they use the app through the number of posts per day.
Due to the one-year post-treatment observation period we naturally exclude users that joined the social network in the final year of the dataset.

\begin{figure}[t]
\centering
\includegraphics[width=.99\columnwidth]{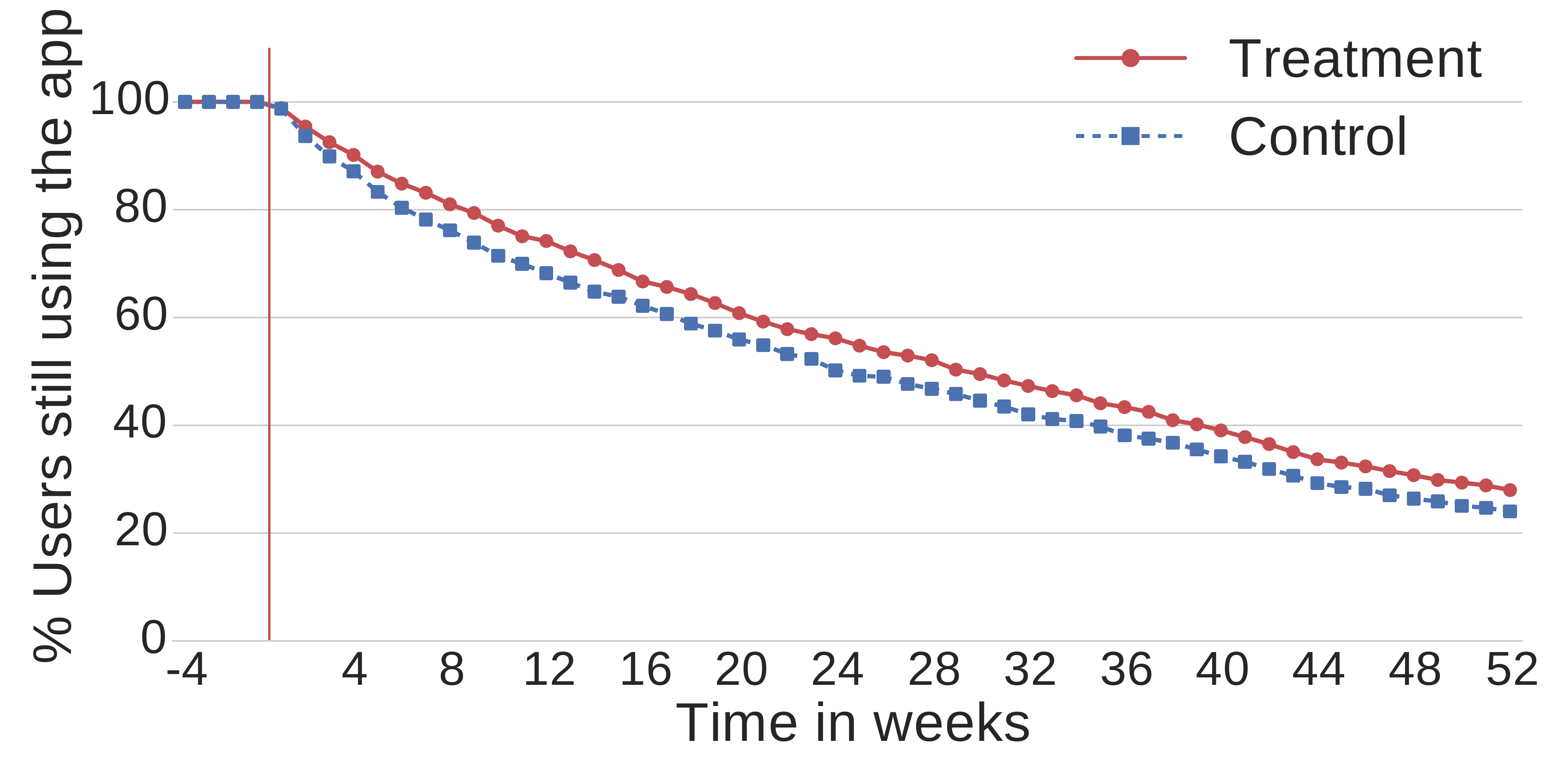}
\caption{
Retention of users that do join the social network at time zero (treatment; red) and matched users that do not (control; blue). 
Social network users are significantly more likely to keep using the activity tracking app during any of the following 52 weeks.
Confidence intervals are too small to be visible.
}
\label{fig:joining_network_retention}

\centering
\includegraphics[width=.99\columnwidth]{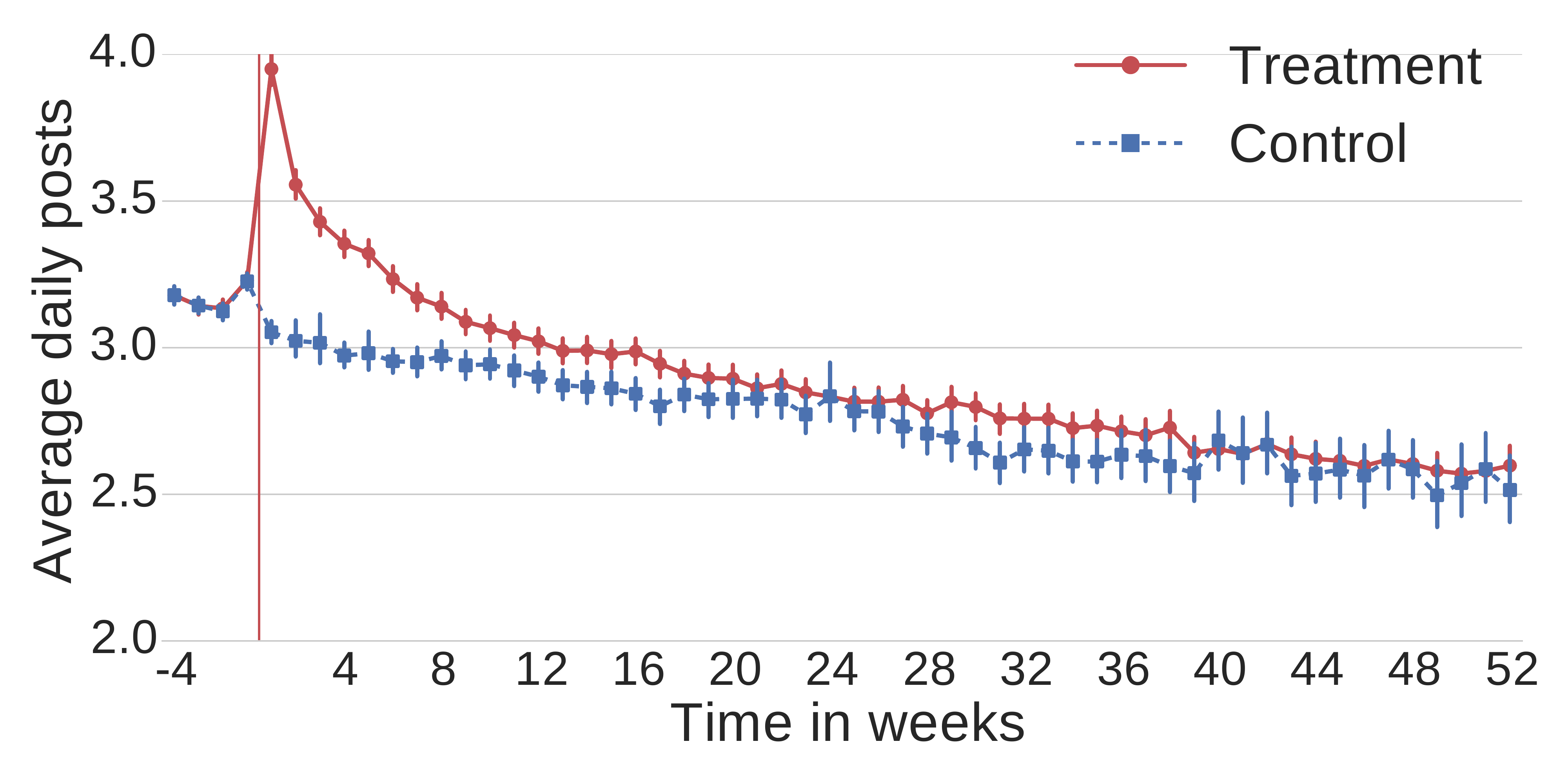}
\vspace{-2mm}
\caption{
App usage of users that do join the social network at time zero (treatment; red) and matched users that do not (control; blue) 
among users still using the app in each week.
Social network users create more posts than control users for a period of about 20 weeks after joining the social network.
}
\label{fig:joining_network_retention_checkins}
\end{figure}

\xhdr{Results}
Retention rates for 6298 matched user pairs are shown in Figure~\ref{fig:joining_network_retention}.
Again, these users are only slightly more active (633 more daily steps) and slightly older (3 years) than the average social network user due to the constraints.
Initially, both groups have 100\% retention rate by design.
After users in the treatment group join the social network, we observe a significant difference in retention between the treatment and control group that becomes clearly noticeable after about three weeks.
This long-lasting effect persists even after one year,
only 24.0\% of control users still use the app whereas 28.0\% of treatment users still do 
($p<10^{-4}$; two-tailed, Mann--Whitney U test). 
This is a 17\% increase in user retention after one year for social network users compared to matched control users.

App usage statistics for the same set of matched user pairs are shown in Figure~\ref{fig:joining_network_retention_checkins}.
Note that only active users with at least one post in each respective week are considered.
Therefore, differences in overall retention rate do not impact these results.
The treatment (red) and control users (blue) have identical levels of average pre-treatment engagement during the four weeks prior to treatment.
As in Section~\ref{subsec:joining_steps}, we note a slight increase in the number of posts one week prior to treatment.
%
Right after joining the social network, treatment users exhibit a large 30\% increase in daily number of posts while app usage in control users stays roughly the same (3.95 vs 3.05 average posts per day; $p<10^{-137}$ according to two-tailed, Mann--Whitney U test). 
Again, we observe a long-lasting effect of creating a first social network connection (and possibly more after that) which exists for over four months.
This effect diminishes over time and the end of the one year long observation period the groups are statistically indistinguishable again (2.59 vs 2.51 average daily posts; $p=0.058$ according to two-tailed, Mann--Whitney U test).

\section[The Effect of Individual Edge Formations]{The Effect of Individual Edge\\Formations}
\label{sec:edge_effects}

\begin{figure*}[t!]
\centering
\includegraphics[width=.49\textwidth]{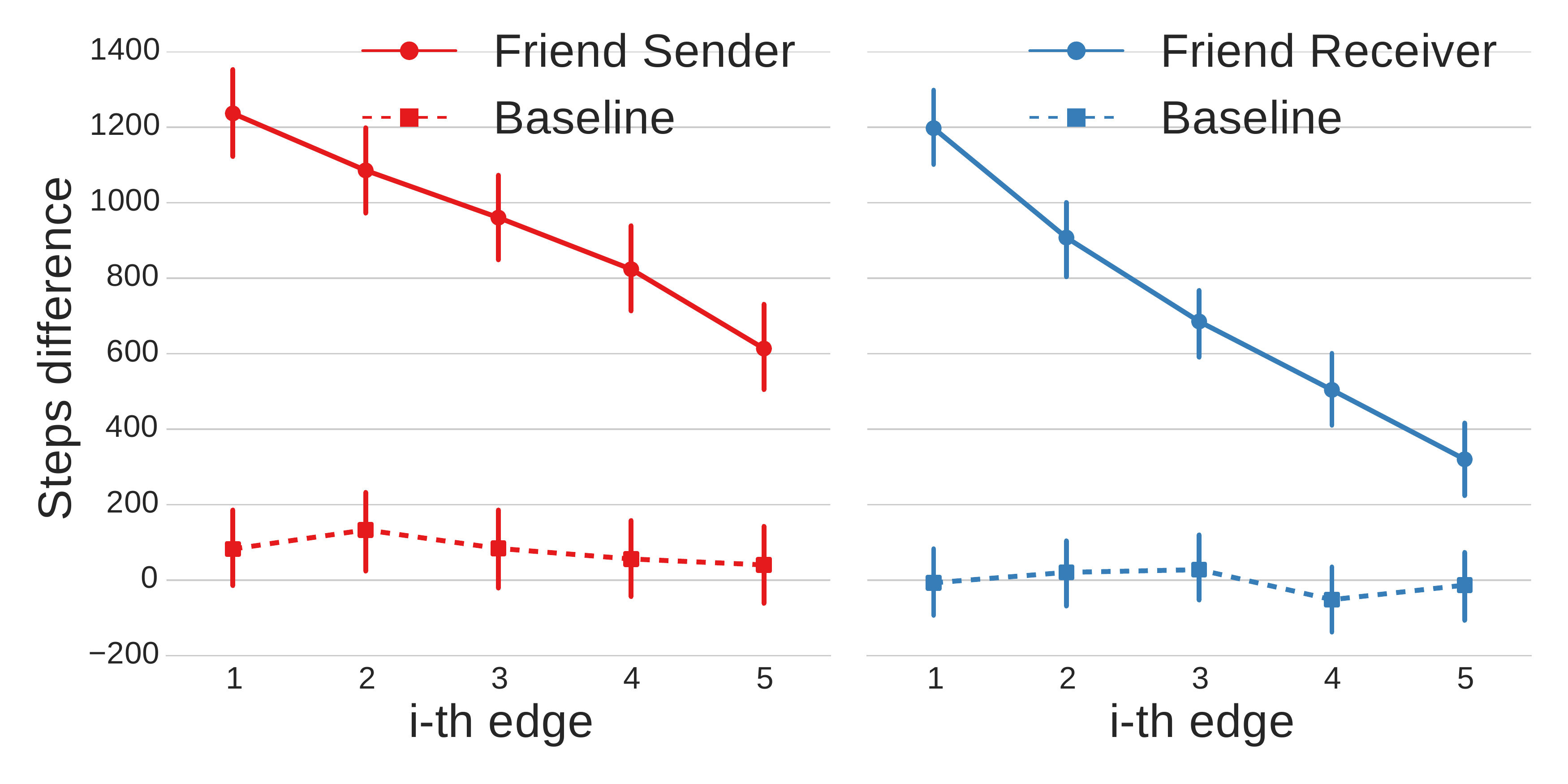}
\includegraphics[width=.49\textwidth]{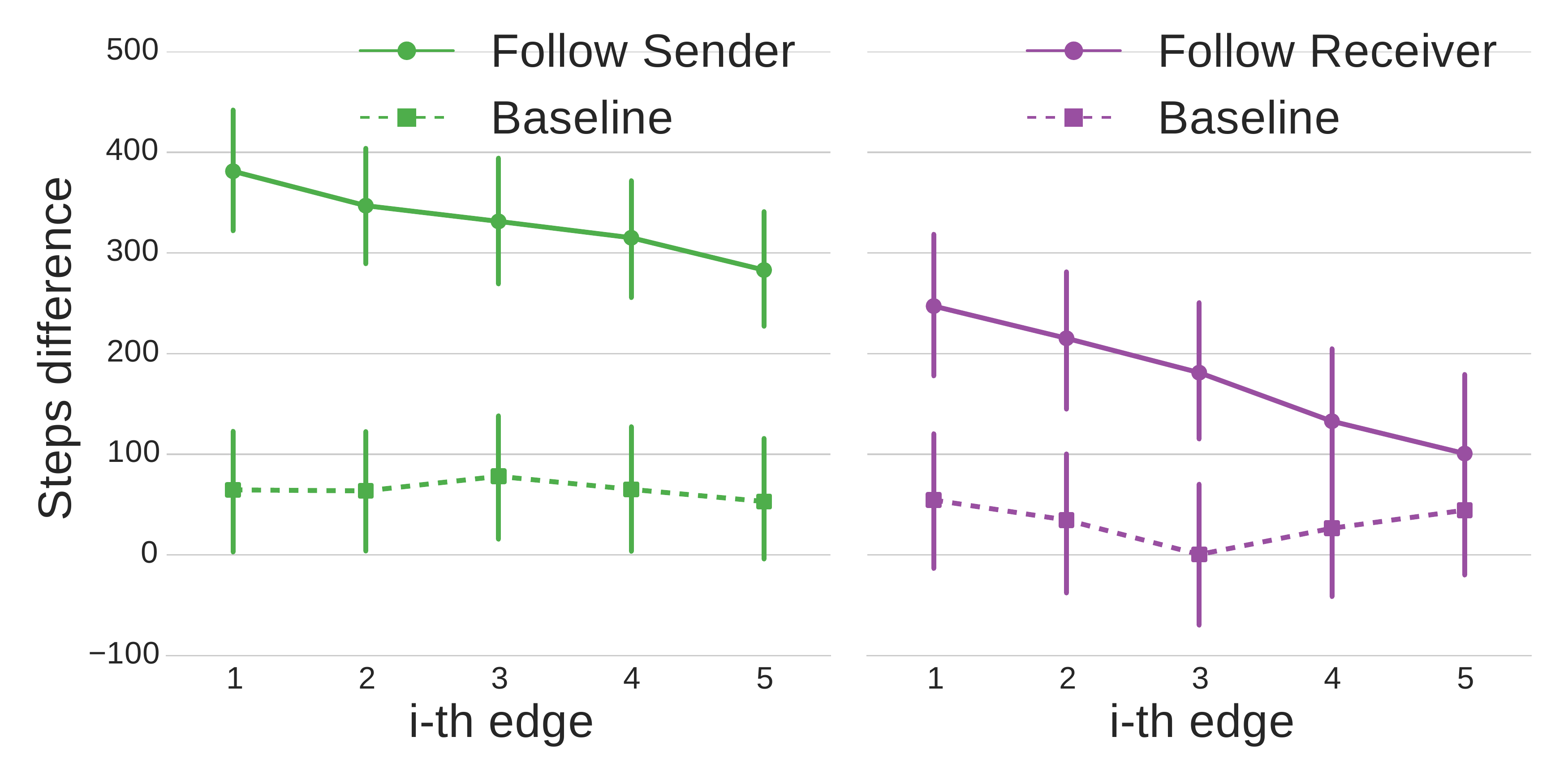}
\vspace{-2mm}
\caption{
The average daily difference in steps 7 days before and 7 days after edge creation as a function of current node degree (x-axis), edge initiator (sender vs receiver), and edge type (friend vs follow).
Dashed lines show corresponding baselines (see Section~\ref{sec:edge_effects}).
We observe significant physical activity increases after edges get created, 
but decreasing effect sizes with each additional edge. 
The effect is larger for edge senders compared to receivers and larger for friend edges compared to follow edges.
}
\label{fig:edge_effects_steps_wbaseline}
\end{figure*}

\begin{figure*}[htp]
\includegraphics[width=.49\textwidth]{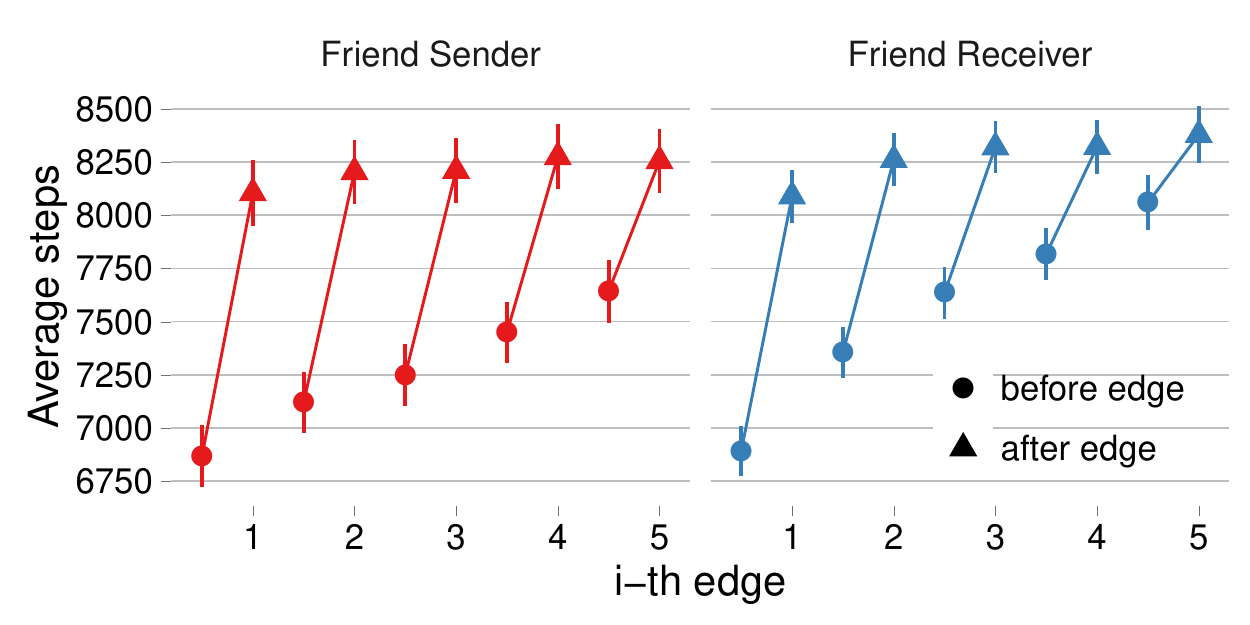}
\includegraphics[width=.49\textwidth]{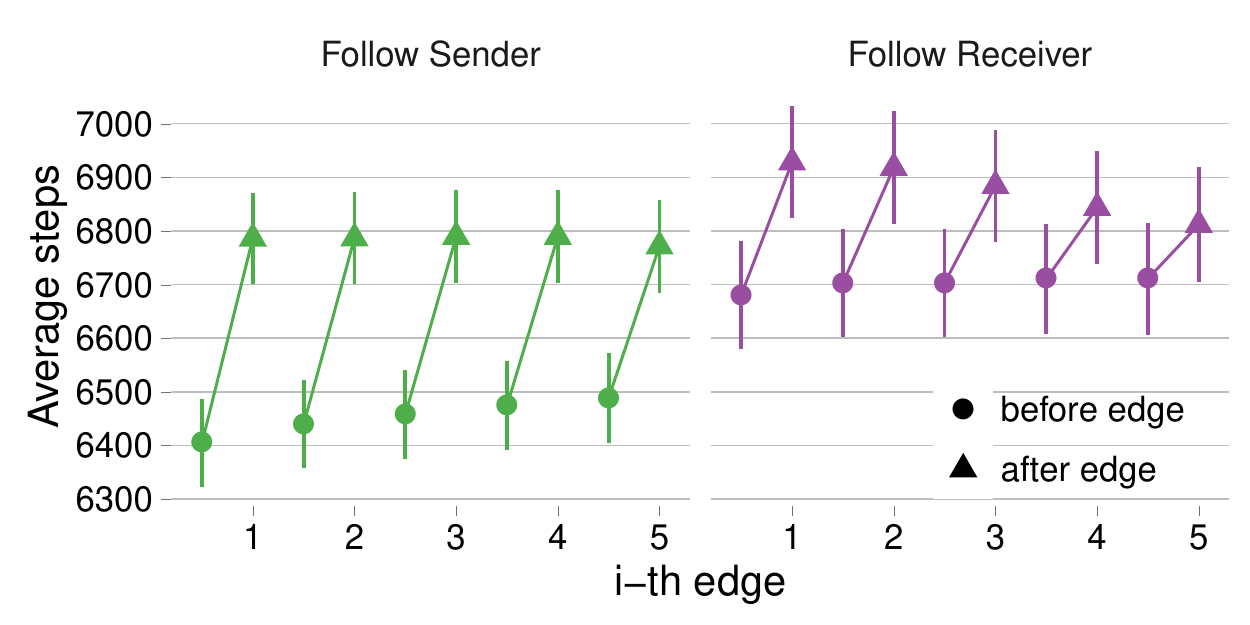}
\vspace{-2mm}
\caption{
Steps before (circles) and after edge creation (triangles) as a function of current node degree (x-axis), edge initiator (sender vs receiver), and edge type (friend vs follow).
We observe increasing activity activity levels before edge creation (circles) while the activity after edge creation varies less (triangles).
This forms an oscillation pattern combined with decreasing effect size (smaller steps differences) at each edge (see Section~\ref{sec:edge_effects}).
}
\label{fig:edge_effects_steps_oscillation}
\end{figure*}

While in Section~\ref{sec:joining_network} we studied the effect of joining the social network through a user's \textit{very first} created edge, 
in this section we measure the effect of each \textit{additional} edge formation; that is the user's second edge, third edge, and so on.
We further distinguish between the type of the edge (friend or follow), the sender and receiver of the edge request, and user demographics (age, gender, weight, and prior activity level).
We show that individual edge formations in the social network are linked to large temporary average increases in daily steps, 
and that the effect decreases with each additional edge.
The effect on user behavior is larger for friend than follower connections, larger for the sender compared to the receiver of the request, and it further varies based on user demographics.

\xhdr{Method}
For every edge that gets created in the social network we measure physical activity right before and right after edge creation (again, we report results for a 7 day window but results are similar for window sizes of 3, 5, and 14 days).
Statistically significant differences in activity before and after are attributed to the edge creation event.
We filter out edges where the request and acceptance for the edge are further than one day apart to reduce bias in the short-term window (10.91\% of edges filtered; see Section~\ref{sec:intrinsic_motivation}).

First, we study how behavior change varies with edge number (\ie, does adding a first friend make a bigger difference than adding the second?), 
and further distinguish edges based on the initiator of the edge (sender vs receiver) and the edge type (friend vs follower edge).
We only consider users with at least five edges and only consider each of their first 5 edges to restrict ourselves to a constant set of users (results for a different number of edges are qualitatively similar). 
Second, we estimate who might be most susceptible to change by estimating the effect for users of different age, BMI, gender, and prior physical activity level.

It is possible that a user's activity in the app changes over time, irrespective of the edge creation event. To account for this we consider the following baseline.
For each user contributing to our estimate, we randomly sample a baseline user with the same app join date ($\pm$ 7 days) and having some recorded activity within 7 days of edge creation to ensure that this user is still active.
We then measure the same difference in activity at the exact same time for the baseline user.
This baseline process is repeated for each combination of edge number (1-5), edge initiator (sender/receiver), and edge type (friend/follow).

Note that due to the small window size of 7 days we only estimate short-term effects.
This is intentional as many factors (including other edges created) could confound the effect estimates for longer-term effects due to the observational nature of the data.

\xhdr{Results}
The estimated effects of each edge on average daily steps are shown in Figure~\ref{fig:edge_effects_steps_wbaseline}, split by edge number (x-axis), edge initiator (sender vs receiver), and edge type (friend vs follow).
Baselines are shown with dashed lines and rectangles.
We observe significant increases in activity after edges are created
up to 1236 average daily steps for the first edge,  
and we observe smaller increases with each additional edge.
Note that the this effect size of up to 1236 additional daily steps is larger than the ones reported reported in Section~\ref{sec:intrinsic_motivation} and Section~\ref{sec:joining_network} for three reasons: 
(1) To show the declining effect with each edge we report the effect in users that have at least five edges and these users are more active than the average social network user.
Measuring the same difference in all social network users gives an average effect estimate of 410 additional daily steps in consistent agreement with previous estimates in Section~\ref{sec:intrinsic_motivation} and Section~\ref{sec:joining_network}.
(2) We differentiate between friend and follower edges.
The effects are between 224\% and 117\% larger for friend edges than for follower edges (1236 vs 381 and 613 vs 283 daily steps; for first and fifth edge for sender, respectively).
This is consistent with the expectation that stronger ties would have a higher potential for social influence~\cite{Aral2014}. 
(3) We differentiate between the sender and receiver of each edge.
Comparing the effect on the sender and receiver,
we find a 3\% larger effect for the sender for the first edge (1236 vs 1197 daily steps)
and a 92\% larger effect for the fifth edge (613 vs 320 daily steps).
All measured behavior changes are significantly larger than their respective baselines (dashed) which are all close to zero change.
We observe very similar results when using tracking behavior (\#days tracked each week) as the outcome variable (online appendix~\cite{althoff2016onlineappendix}).

\xhdr{Explaining decreasing influence of edges}
Figure~\ref{fig:edge_effects_steps_oscillation} shows average daily steps before (circles) and after edge creation (triangles), again split by edge number (x-axis), edge initiator (sender vs receiver), and edge type (friend vs follow).
Across node degree, activity shows an oscillation pattern explaining the decreasing influence effects observed in Figure~\ref{fig:edge_effects_steps_wbaseline}. 
After the boost of the first edge (compare first triangle and circle), the activity just before the creation of the second edge drops down again but to a level higher than before the first edge was created.
The next edge then comes with another slightly smaller boost in activity compared to the first edge. 
This shows that the decreasing effects are largely due to increasing activity levels before each edge creation.

\begin{figure}[t]
\centering
\includegraphics[width=\columnwidth]{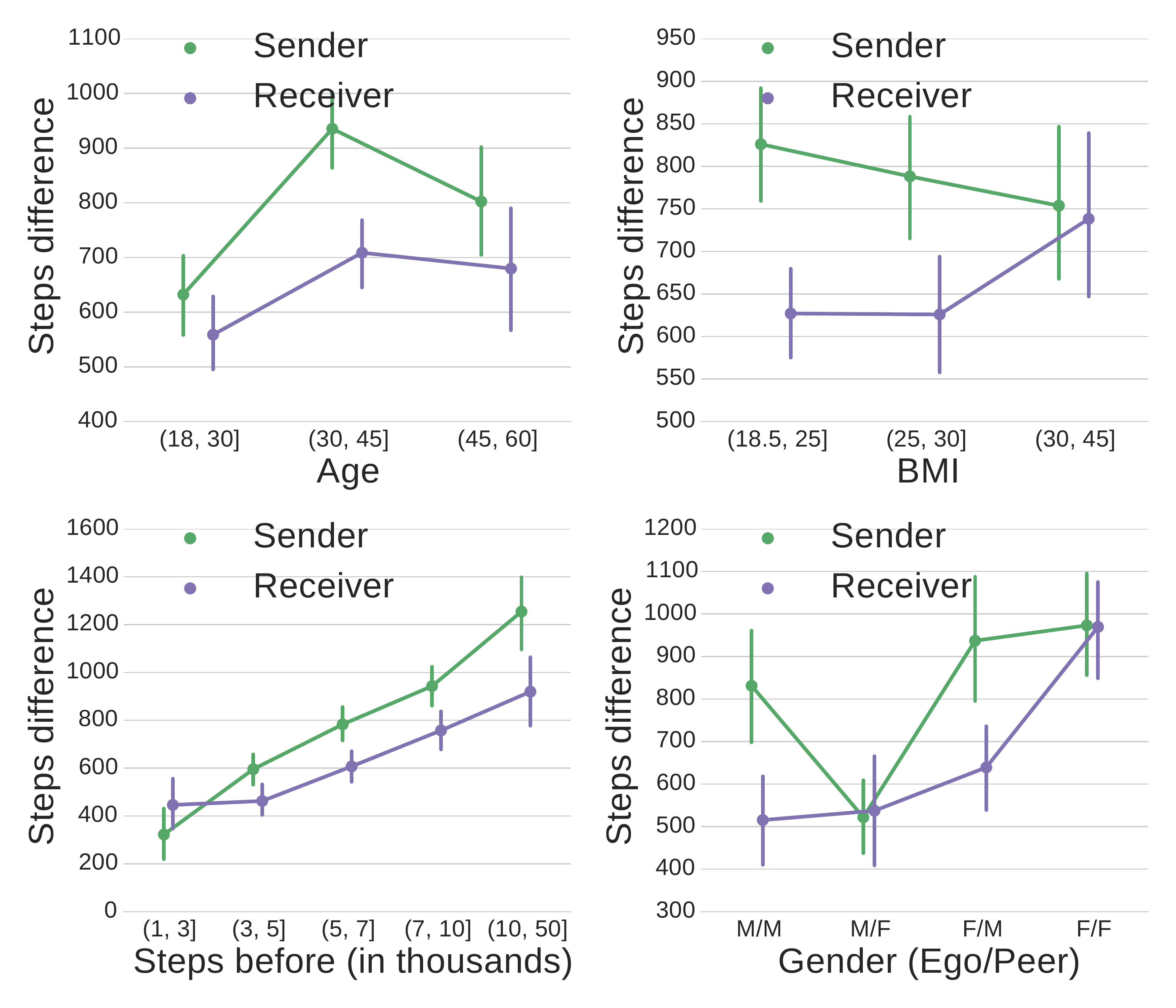}
\vspace{-4mm}
\caption{
The difference of steps after edge creation as a function of the user's age, BMI, prior physical activity level, and the interaction between gender of sender and receiver.
We observe larger changes in behavior for users that are older, have higher BMI, and take more steps in the week before.
}
\label{fig:edge_effects_who_is_most_susceptible}
\end{figure}

\xhdrnodot{Which users are most susceptible to change?}
Previous work has reported mixed results on differences in behavior change across age and gender. 
Dishman and Buckworth~\cite{dishman1996increasing} find no such differences in the context of physical activity.
However, a study on product adoption on Facebook found that susceptibility decreased with age and was slightly lower in women~\cite{Aral2012susceptible}.  
As before, we consider the first 5 friend edges for all users with at least 5 friend edges.
We estimate the difference in steps for users of different age, BMI, physical activity level, and gender.
The results are shown in Figure~\ref{fig:edge_effects_who_is_most_susceptible}.
We find larger changes for older users, with the largest changes for 30-40 year olds (935 steps for edge sender),
which is noteworthy since physical activity typically decreases with age~\cite{Bauman2002}.
For edge senders, we find slightly smaller steps differences for users that are overweight (25<=BMI<30) or obese (BMI>=30) which is consistent with prior findings that healthy individuals typically exhibit larger behavioral differences~\cite{dishman1996increasing}.
Surprisingly, the pattern is reversed when receiving an edge and obese individuals change 18\% more than normal weight individuals.
This suggests that offering a friendship to an obese person 
is particularly beneficial to them.
Further, we find that people with more steps are most susceptible to larger changes with averages increases of 1254 daily steps (for sender) for users that typically take ten thousand or more daily steps on average.
This is consistent with existing theories that healthy individuals are capable of larger behavior change~\cite{dishman1996increasing}.
Lastly, we find a significant interaction between the ego and the peer's gender for behavior change.
When men send a friend request to a woman, they change their behavior by 37\% less compared to when they send it to another man.
Women seem to particularly appreciate receiving friend requests from other women and increase their step count 52\% more compared to when receiving a friend request from a man.
In all cases, the sender of the edge changes at least as much as the receiver of the edge even though friend connections are bi-directional. 

\section{Predicting Behavior Change}
\label{sec:prediction}

In the previous sections we studied the average aggregate influence of edges in a social network.  
This section summarizes the insights obtained in this paper in a series of predictive models to predict which individual users will be most influenced by the creation of new social network connections on the level of individual edges.
We demonstrate that while there is large variability in users' responses to new friends and followers, 
the factors described in this work allow to predict user behavior change with significant accuracy.

\xhdr{Method}
We formulate the prediction task as a binary prediction of whether or not someone increases or decreases their activity in the 7 day window after adding a new friend or follower (either as sender or receiver; 55.4\% of the time users increase their activity compared to the previous 7 days).
We use the dataset of all 432,133 edge creations for which we can measure change in physical activity 
(\ie, we observe steps both before and after the edge gets created).
Prediction is performed on both the full dataset as well as the subsets of the data split by edge type (friend/follow) and initiator (sender/receiver).
In all cases, we use a balanced dataset by randomly subsampling the majority class
and use 80\% for training and 20\% for testing.
Area under the ROC curve is used as a measure of predictive performance on the test set.
We report performance for Gradient Boosted Tree models and optimize number of trees, tree depth, and learning rate through cross-validation on the training data.
We also experimented with Logistic Regression and linear SVM which consistently gave lower performance, especially when feature interactions appeared useful (\eg, demographics).

\xhdr{Models}
We define a series of models with increasing complexity in order to learn what features are most useful in prediction of behavior change.
If features are missing we impute zero and include a binary variable indicating missingness.
\begin{enumerate}
\item Random Baseline: Included for comparison.
\item Previous behavior change: Activity increase or decrease in steps after the most recent edge creation of the same type and initiator (note that this is not available for anyone's first edge). We only use previous edges that were created at least 7 days prior to the current edge because otherwise this feature could give away the true label for the current edge.
\item Edge type (friend vs follow), edge initiator (sender vs receiver) and edge number
\item User demographics: Age, gender, and BMI.
\item Steps before: Average number of steps in the 7 day window before edge creation.
\item All features: Combination of models 2-5.
\end{enumerate}
\vspace{-2mm}
\xhdr{Results} 
Prediction accuracies are shown in Table~\ref{tab:prediction}.
We first focus on the prediction accuracies on the full dataset (first column).
Knowing how an individual responded to previous edge creations is not very useful for predicting the response to the current edge (AUC=0.538).
This shows that there is significant variation over time even within a single user.
As shown in Section~\ref{sec:edge_effects}, behavior change varies with edge type, initiator and edge number and using only these features gives an AUC of 0.574.
Demographic features (age, gender, BMI) are very predictive of how an individual might respond after edge creation (AUC=0.685).
However, the strongest single feature is the physical activity level before edge creation (AUC=0.715).
Combining all features (2-5) gives the highest predictive performance at an AUC of 0.785.

For the individual edge types and initiators (last four columns in Table~\ref{tab:prediction}), 
we find that behavior change after friend edges is significantly less predictable than after follow edges.
This is largely due to the fact that ego demographics (model 4) and previous activity level (model 5) are both particularly good predictors for follow edges.
Further, the performance is higher for predicting change in the sender than the receiver.
 
Overall, we demonstrate that we can predict who will be influenced to increase their activity after edge creation 
with our proposed models achieving an AUC of over 78\%.

\begin{table}[t]
\centering
\small 
\resizebox{.999\columnwidth}{!}{%
\begin{tabular}{rp{22mm}rrrrr}
  \toprule
  & \textbf{Model} &  \textbf{All} & \textbf{Fr/S} & \textbf{Fr/R} & \textbf{Fo/S} & \textbf{Fo/R} \\ 
  \midrule
  1 & None   &  0.500 &  0.500  &   0.500  &  0.500  &  0.500\\ 
  2 & Previous behavior change &  0.538 &  0.543  &   0.551 &  0.546  &  0.526\\
  3 & Edge type, initiator and number &  0.574 &  0.515  &  0.518 &  0.506  &  0.510\\
  4 & User demographics (age, gender, BMI) &  0.685 &  0.644  &   0.583 &  0.777  &  0.773\\
  5 & Steps 7 days before &  0.715 &  0.665  &  0.614 &  0.808  &  0.781\\
  6 & All features (2-5) &  \textbf{0.785} &  \textbf{0.721}  &   \textbf{0.672} &  \textbf{0.847}  &  \textbf{0.830}\\
  \bottomrule
 \end{tabular}
 }
 \caption{
 Performance of several models predicting activity increase or decrease after edge creation using the area under the ROC curve.
 The table reports predictive performance on all data (all), and split by edge type and initiator: Friend (Fr), Follow (Fo), Sender (S), Receiver (R).
 }
 \label{tab:prediction}
 \end{table}

\section{Related Work}
\label{sec:related_work}

More recently, there has been a lot of interest in using 
social media and online social network datasets as a large-scale sensor into people's activity and health. Studies examined diet and weight loss~\cite{de2016characterizing,ma2010analysis,mejova2015foodporn}, 
mental health~\cite{de2016discovering}, 
and exercise~\cite{ebrahimi2016characterizing}. 
Other work has used long-term medical studies to study obesity~\cite{Christakis2007obesity} and smoking~\cite{Christakis2008smoking}, smart phone and wearable device data to study health benefits of physical activity~\cite{althoff2016quantifying,althoff2016pokemon}, and text messaging data to study mental health~\cite{althoff2016counseling}.
Research on using online social networks for health interventions is still its infancy~\cite{maher2014health} and mixed findings have been reported on its effectiveness to encourage physical activity~\cite{cavallo2012social,foster2010motivating}. 
For example, in a user study to encourage physical activity with 13 participants all but one reported to be motivated by social influence~\cite{consolvo2006design}.

Our work extends prior research in the following key aspects:
First, we study offline user behavior through physical activity, a behavior critical to human health (\eg, physical inactivity contributes to 5 million deaths every year~\cite{Lee2012pa}).
Second, we use a comprehensive dataset capturing the 
introduction and growth of the network and changes in user behavior, 
allowing us to identify influence effects exactly when new connections are made.
With 6 million users recording 631 million activity posts and 160 million days of steps tracking over the course of 5 years, our study is substantially larger than comparable studies on social network effects on physical activity (254 users and 265 posts in \cite{ebrahimi2016characterizing}).
Third, we employ natural experiments~\cite{rosenzweig2000natural}, difference-in-difference designs~\cite{Lechner2011diffindiff}, and matching procedures~\cite{rosenbaum2010design,stuart2010matching} from the econometrics literature to identify causal relationships from observational data.
Fourth, while much online social media research 
has relied on self-reported behaviors and proxy measures (\eg, \cite{althoff2014howtoaskforafavor,althoff2015donor,de2016discovering,de2016characterizing,mejova2015foodporn}), 
we use objective measures of physical activity through smart phone accelerometers (\eg, \cite{althoff2016quantifying,althoff2016pokemon}).
This is critical as self-reports of any behavior can be biased~\cite{belli1999reducing,hebert1995social} 
and in particular self-reports of physical activity have been found to be over-estimates by up to 700\%~\cite{Tucker2011selfreport}.

\section{Conclusion}
\label{sec:conclusion}

We present several natural experiments and observational studies about how online social network features influence offline and online user behavior using a large activity tracking dataset.
Exploiting a natural experiment in delayed friendship formation, 
we distinguish the effect of social influence of single social network connection from simultaneous intrinsic motivation and show that social influence accounts for 55\% of the observed effects.
We show that joining a social network significantly increases  user online in-application activity, user retention, and user offline real-world physical activity.
Further, we demonstrate that each individual edge formation in the social network is linked to temporary average increases in daily steps.
Lastly, we demonstrate that based on the insights obtained in this work, susceptibility to behavior change can be predicted with significant accuracy.
Our work opens up several exciting directions for future work including
studying what incentives lead to healthy behavior change in different groups of people,  
and how to design effective, contextual, and potentially changing interventions.

\xhdr{Acknowledgments}
We thank Bojan Bostjan\v{c}i\v{c} and Peter Kuhar for donating the data for independent research, 
and Hima Lakkaraju, David Hallac, Jen Hicks, Rok Sosi\v{c}, Will Hamilton, David Jurgens, and the anonymous reviewers for their valuable feedback on the manuscript.
This research has been supported in part by NSF
IIS-1149837,
NIH BD2K Mobilize Center,     
ARO MURI, DARPA SIMPLEX and NGS2,
Stanford Data Science Initiative,
Lightspeed, SAP,
and Tencent.

\balance
\bibliographystyle{abbrv}
{\normalsize
\bibliography{refs}}

\pagebreak
\section{Appendix}
\label{sec:appendix}

\subsection{Degree Distribution of Social Network}

\begin{figure}[h!]
\centering
\includegraphics[width=.49\columnwidth]{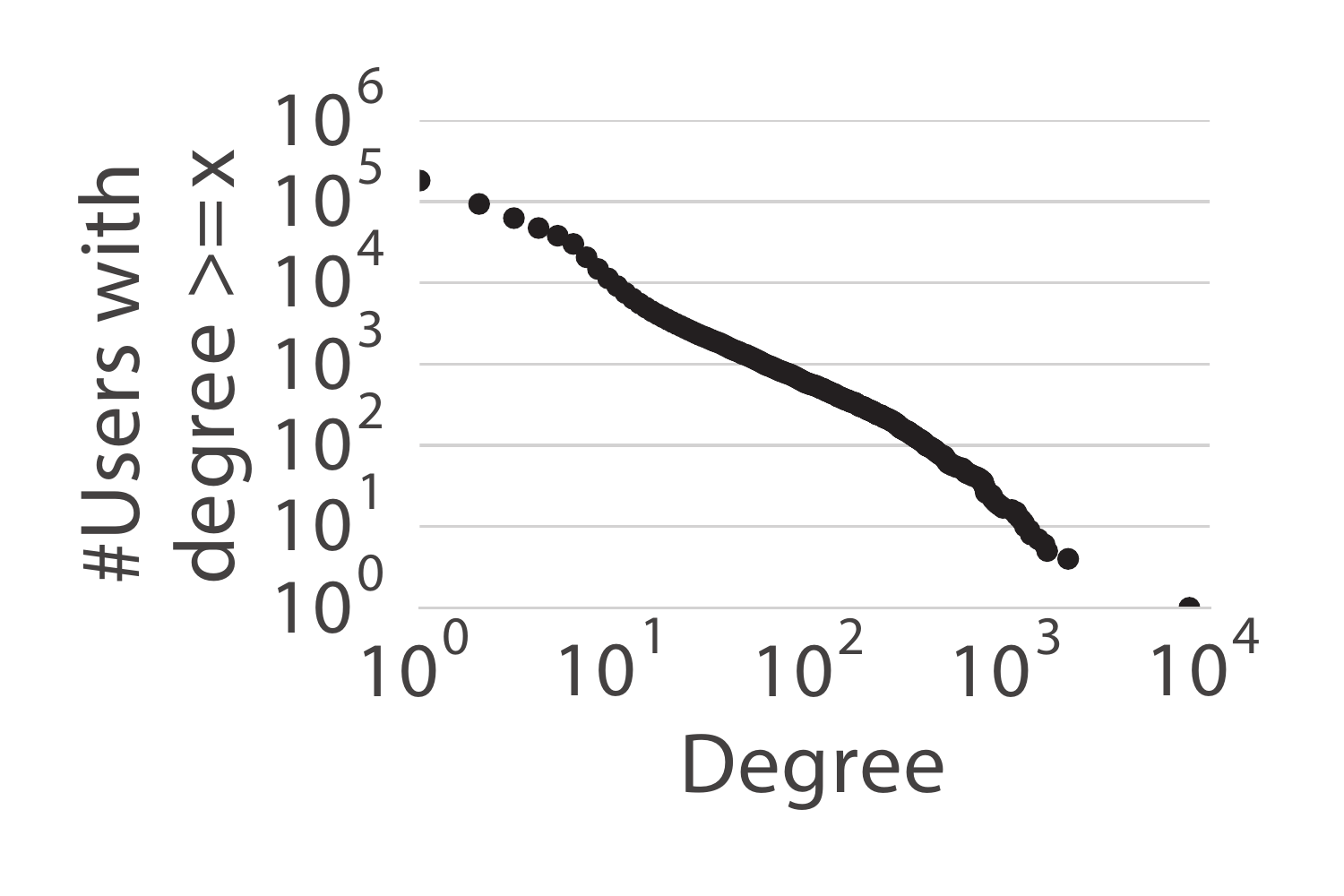}
\includegraphics[width=.49\columnwidth]{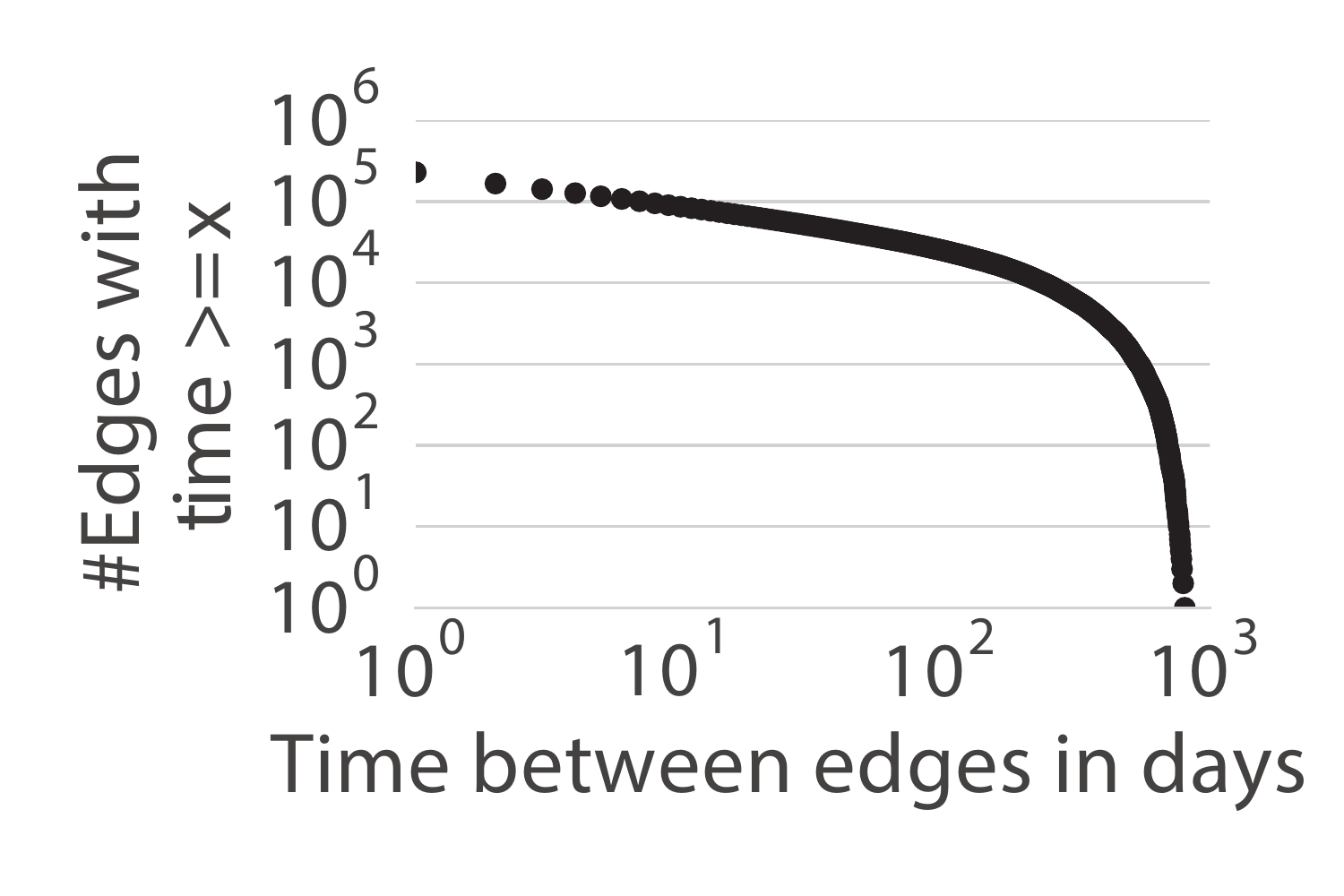}
\caption{Left: Degree distribution for social-network users including both bi-directional friend and uni-directional follower connections.
Right: Distribution of time between edges.
84\% of all edges get created within 7 days of the previous edge.
Both plots show unnormalized CCDFs. 
}
\label{fig:degree_distribution_and_time_between_edges_distribution}
\end{figure}

\subsection{Effect of Joining the Social Network: Additional Short-term Experiment}
\begin{figure}[htp]
\centering
\includegraphics[width=\columnwidth]{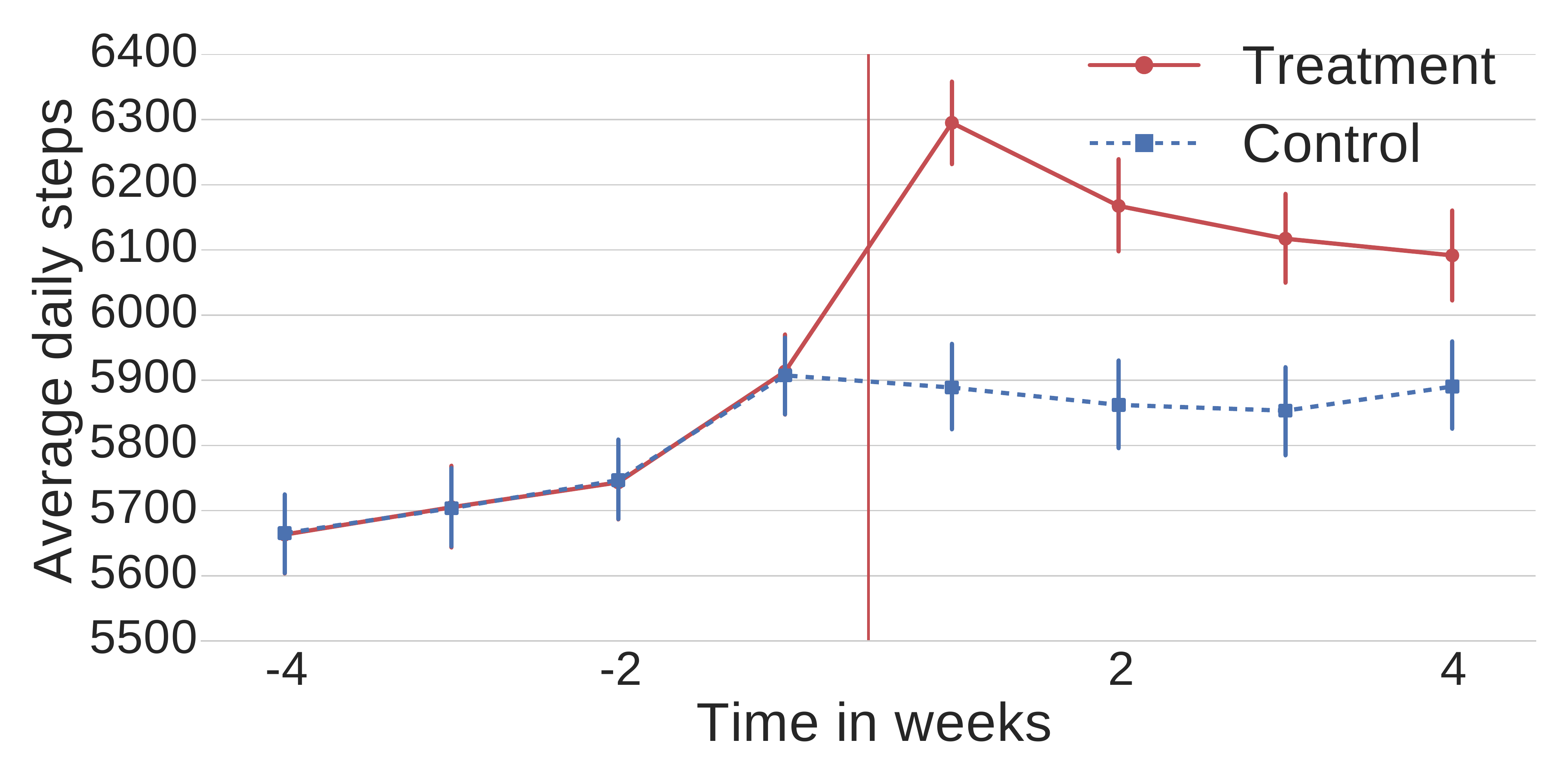}
\caption{Average daily steps for users that do join the social network at time zero (treatment; red) and matched users that do not (control; blue). 
We observe a significant boost in activity in treatment users but no difference in control users.
The effect size is practically identical with Figure~\ref{fig:joining_network_design4}.
}
\label{fig:joining_network_design4_short_term}
\end{figure}

We repeat the experiment described in Section~\ref{subsec:joining_steps} with less stringent constraints where we require users to track steps for only 4 weeks after joining the social network. 
After joining the social network, treatment users take 382 daily steps more on average (Figure~\ref{fig:joining_network_design4_short_term}).
This effect size for this weaker constraint is practically identical to earlier estimates (Section~\ref{sec:intrinsic_motivation} and Section~\ref{sec:joining_network}).

\subsection{Distinguishing between Intrinsic Motivation and Social Influence: Additional experiment using Pending Requests}
\label{subsec:pending_requests}

A central finding in our paper is a causal effect of how social networks influence user behavior.
To provide further evidence for this effect, we additionally repeat the analysis of Section~\ref{sec:intrinsic_motivation} but using \textit{pending} (\ie, never accepted) edge requests instead of delayed accepted ones (\ie, accepted after 7 days or longer).
We show that this alternative analysis leads to very similar effect size estimates providing further validation of our natural experiment methodology in Section~\ref{sec:intrinsic_motivation}.

\subsubsection{Using Pending Friendship Requests to Estimate Social Influence}
\label{subsubsec:using_pending_requests}
\xhdr{Method}
We use the conceptual framework described in Section~\ref{sec:intrinsic_motivation} based on a difference-in-difference analysis~\cite{Lechner2011diffindiff} where we measure the difference in physical activity levels 7 days after compared to 7 days before the edge creation time stamp for both accepted and pending friend requests.
As in Section~\ref{sec:intrinsic_motivation} we only consider edge requests that correspond to friend requests and exclude any requests that do have other requests within a 7 day window around the time of request as these could lead to compounding effects.
We further exclude 4.69\% of edge requests that were accepted later than 24 hours because we want to focus the start of the potential influence effect and not consider edge requests that are accepted much later (note that we studied the effect of these delayed accepted edges in Section~\ref{sec:intrinsic_motivation}).

\xhdr{Results}
The differences in post-treatment activity for both accepted and pending friend requests are shown in Figure~\ref{fig:pending_vs_accepted_steps_matching} (left panel).
These results provide further confirming evidence for the results based on delayed accepted requests presented in Section~\ref{sec:intrinsic_motivation} as we find very similar effect sizes.
In particular, we find again that behavior changes significantly even after pending friend requests (100 daily steps difference; significantly larger than 0 according to Wilcoxon signed rank test; $p<10^{-2}$). 
We  also find that the effect for accepted friend requests is significantly larger (330 daily steps difference; $p<10^{-15}$ Mann--Whitney U test). 
In this case, we estimate that 30\% (100/330) of behavior change is due to the user's elevated intrinsic motivation and 70\% is due to social influence.

\subsubsection{Matching-based Experiment}
\label{subsubsec:matching}
To provide further evidence for the causal social influence effect, we also conduct an experiment where we match similar pending and directly accepted edges to rule out any potential confounding that could stem from a large number observed covariates. 
However, we find that that this additional experiment also confirms the findings presented in the main paper in Section~\ref{sec:intrinsic_motivation} and the additional experiment described in Section~\ref{subsubsec:using_pending_requests}. 
We continue to employ the conceptual framework of Section~\ref{sec:intrinsic_motivation}, but apply it to matched pairs of edge requests.

\xhdr{Matching}
In Section~\ref{sec:intrinsic_motivation} where we studied delayed accepted edge requests, we observed that all covariates described in Table~\ref{tab:natural_experiment_balancing} were very well balanced.
In the case of pending edge requests, the two groups are less well balanced (median absolute SMD = 0.156, maximum absolute SMD = 1.169).
We address these challenges by creating matched pairs of pending and accepted friend requests using all the covariates listed in Table~\ref{tab:natural_experiment_balancing}.
We note that the relationship between the sending and receiving users does not necessarily need to be fundamentally different for pending and accepted edges.
Users have the ability to \textit{ignore} friend requests giving an explicit signal that this relationship is different from the kind of requests they would accept.
We exclude those ignored friend requests for this exact reason.
Therefore, pending requests might be pending simply because the other person happened to stop using the app.
To verify that our results still hold even if pending and accepted edge requests were different across unobserved covariates, we conduct a sensitivity analysis to test the robustness of our claims~\cite{rosenbaum2010design}.

We use one-to-one almost exact matching~\cite{rosenbaum2010design} of accepted edges to pending edge requests as pending edge requests occur less frequently (we obtain similar results when using Mahalanobis Distance Matching and Propensity Score Matching).
This results in 6900 pairs that are very well matched on all covariates (median absolute SMD = 0.006, maximum absolute SMD = 0.168; \ie, all SMDs below 0.25).

\begin{figure}[tp]
\centering
\includegraphics[width=\columnwidth]{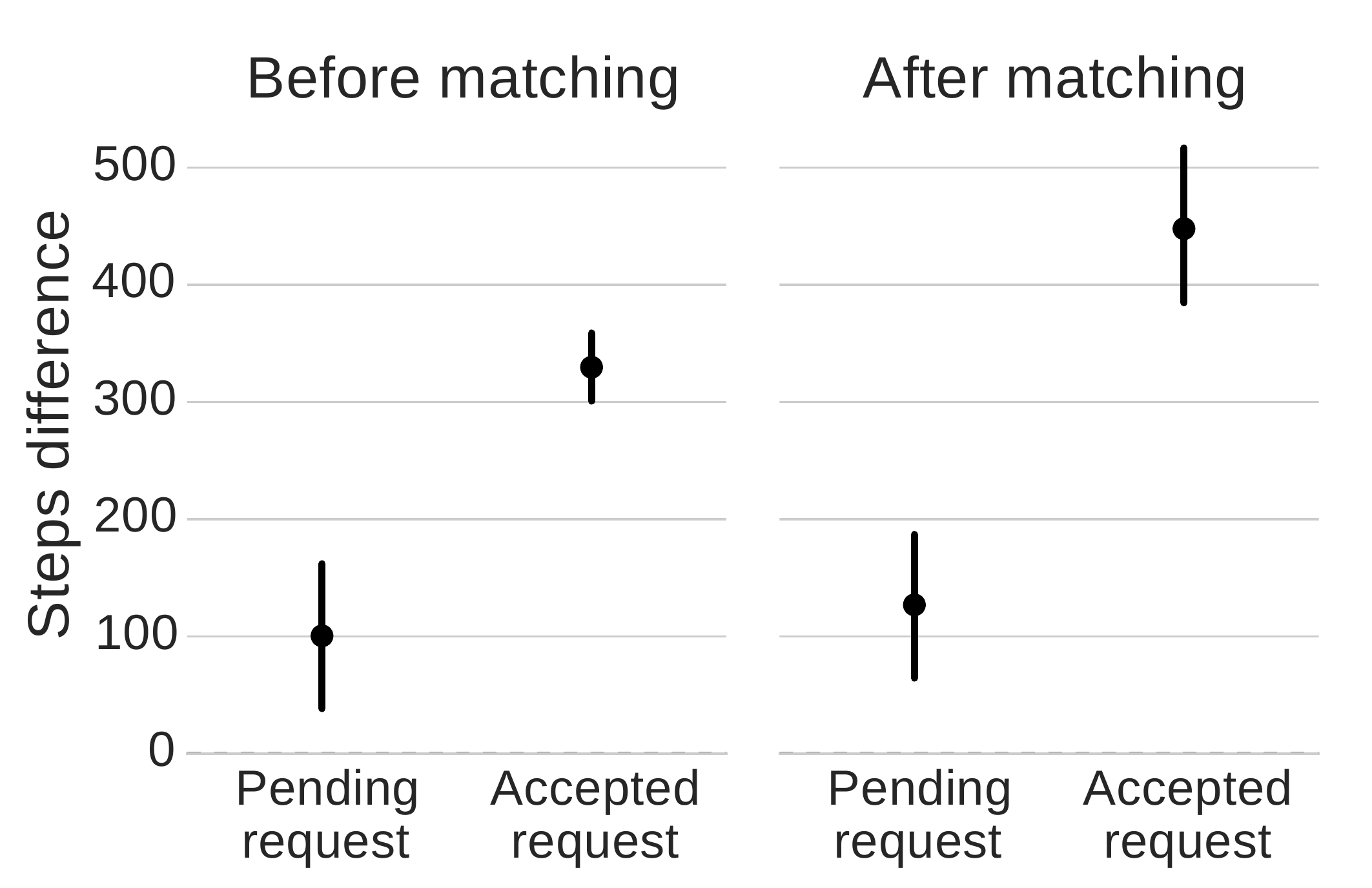}
\caption{
Steps difference after edge creation for accepted and pending edge requests before matching (left) and after matching (right). 
For more information on matching see Section~\ref{subsec:pending_requests}.
Even pending edge requests lead to significant increases in activity.
However, intrinsic motivation (as measured by pending edges) only explains 28\% of the effect observed for accepted edges. 
Therefore, we attribute the remaining 72\% to social influence.
}
\label{fig:pending_vs_accepted_steps_matching}
\end{figure}

\xhdr{Matching results}
The difference in post-treatment activity for both matched groups is shown in Figure~\ref{fig:pending_vs_accepted_steps_matching} (right panel).
The results for the matched groups are very similar to the full data before matching and contributes additional evidence for the social influence effects reported above.
Again, we find that behavior changes significantly even after pending friend requests (127 daily steps difference; significantly larger than 0 according to Wilcoxon signed rank test; $p=0.001$).
We also find that the effect for accepted friend requests is significantly larger (448 daily steps difference; $p<10^{-12}$; 
Wilcoxon signed rank test), even after controlling for a large number of covariates as explained above.
Here, we estimate that 28\% (127/448) of behavior change is due to the user's elevated intrinsic motivation and 72\% is due to social influence.
This is very close to our earlier estimate of 30\% for intrinsic motivation and 70\% for social influence.
Lastly, these results could be invalidated if the two groups differed significantly in terms of some unobserved confounding variable that is causing the effect.
However, since we are controlling for a large number of important covariates (Table~\ref{tab:natural_experiment_balancing}) it is unlikely that there are unobserved that would dramatically change the probability of treatment assignment.
Sensitivity analysis of the results shows that the difference between the two groups is statistically significant (Wilcoxon signed rank test at $p=0.05$) even if the odds of treatment for two matched individuals with identical observed covariates differed by up to a factor of 1.18 (Rosenbaum's $\Gamma=1.18$, see~\cite{rosenbaum2010design}).
This lends further credibility to the robustness of our main finding: Online social networks influence user behavior.

\subsection{Additional Results on the Effect of Individual Edge Formations on Tracking Behavior}
This section extends Section~\ref{sec:edge_effects} which studied the effect of individual edge formations on offline physical activity.
Here, we provide additional results of the effect of individual edge formations on the propensity of users to track their physical activity on a given day (``steps tracking'').

\begin{figure*}[tp]
\centering
\includegraphics[width=.49\textwidth]{edge_effects_baselines_steps_friend.pdf}
\includegraphics[width=.49\textwidth]{edge_effects_baselines_steps_follow.pdf}
\includegraphics[width=.49\textwidth]{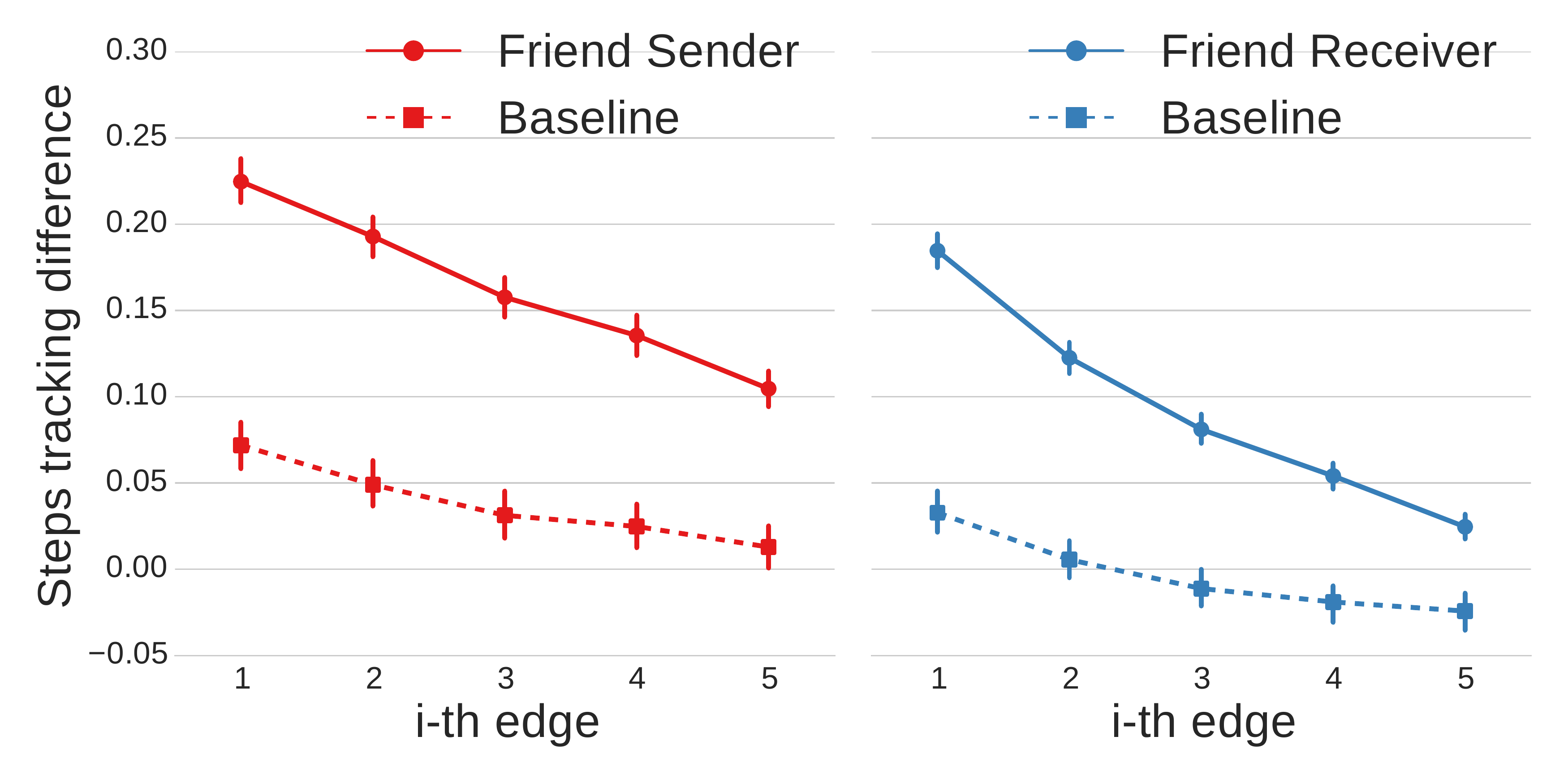}
\includegraphics[width=.49\textwidth]{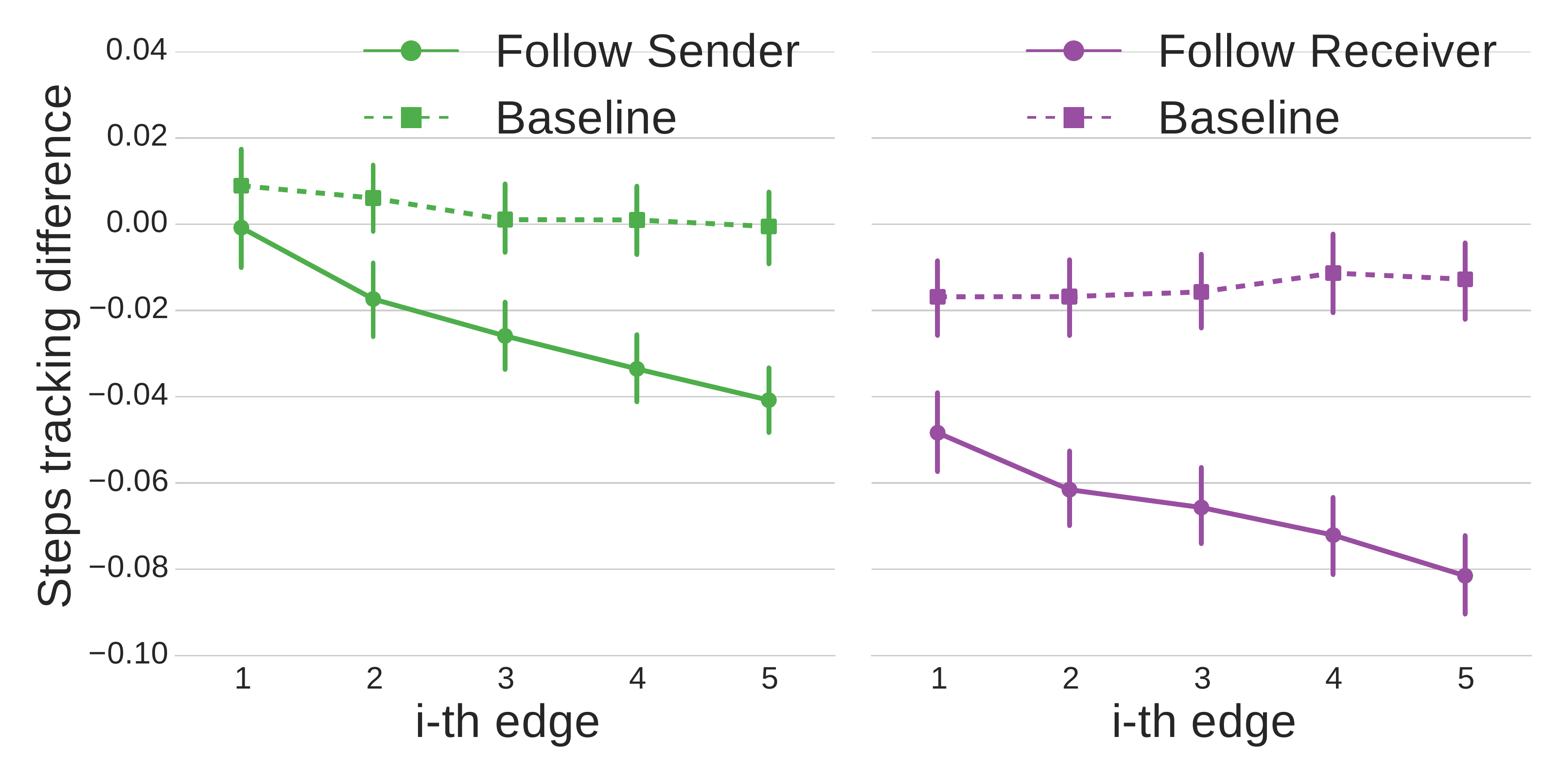}
\caption{The difference in steps (top) and steps tracking (bottom) before and after edge creation as a function of edge number (x-axis) and edge type and initiator (subplots).
Steps tracking difference refers to the difference in probably of recording steps on a given day of the week after edge creation compared to the week before.
Dashed lines show corresponding baselines (main text).
We observe significant activity increases after edges get created. 
The effect is larger for friend edges compared to follow edges and larger for edge senders compared to receivers.
}
\label{fig:edge_effects_steps_wbaseline_appendix}
\end{figure*}

\begin{figure*}[tp]
\centering
\includegraphics[width=.49\textwidth]{oscillating_steps_friends_nozag.pdf}
\includegraphics[width=.49\textwidth]{oscillating_steps_follow_nozag.pdf}
\includegraphics[width=.49\textwidth]{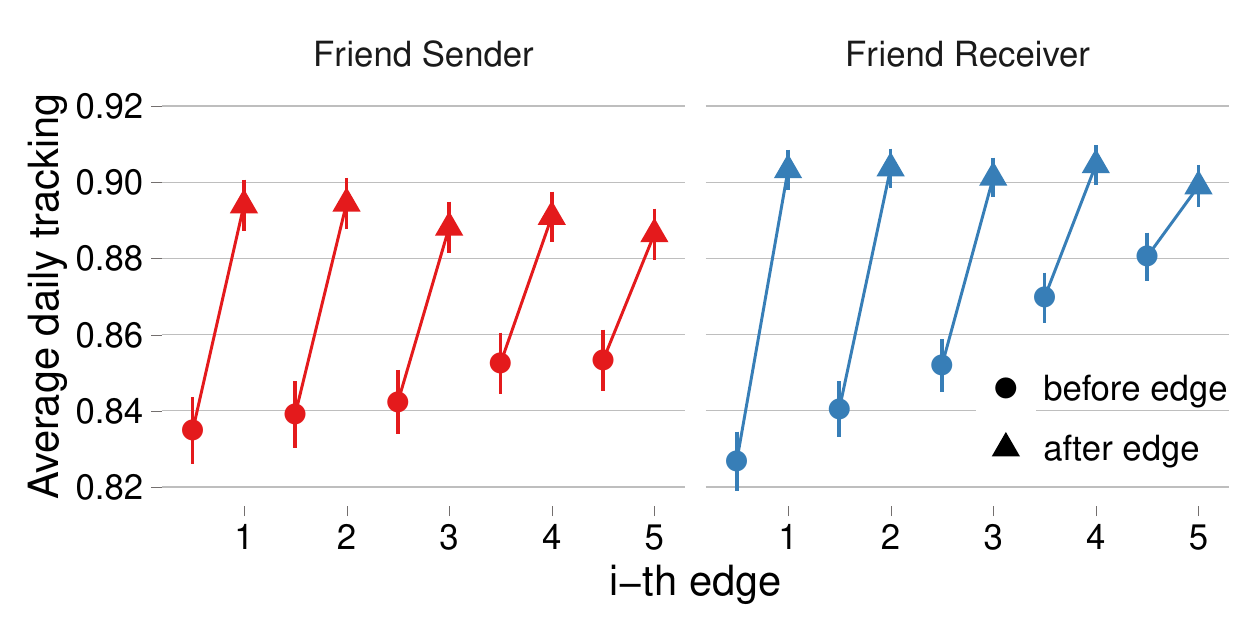}
\includegraphics[width=.49\textwidth]{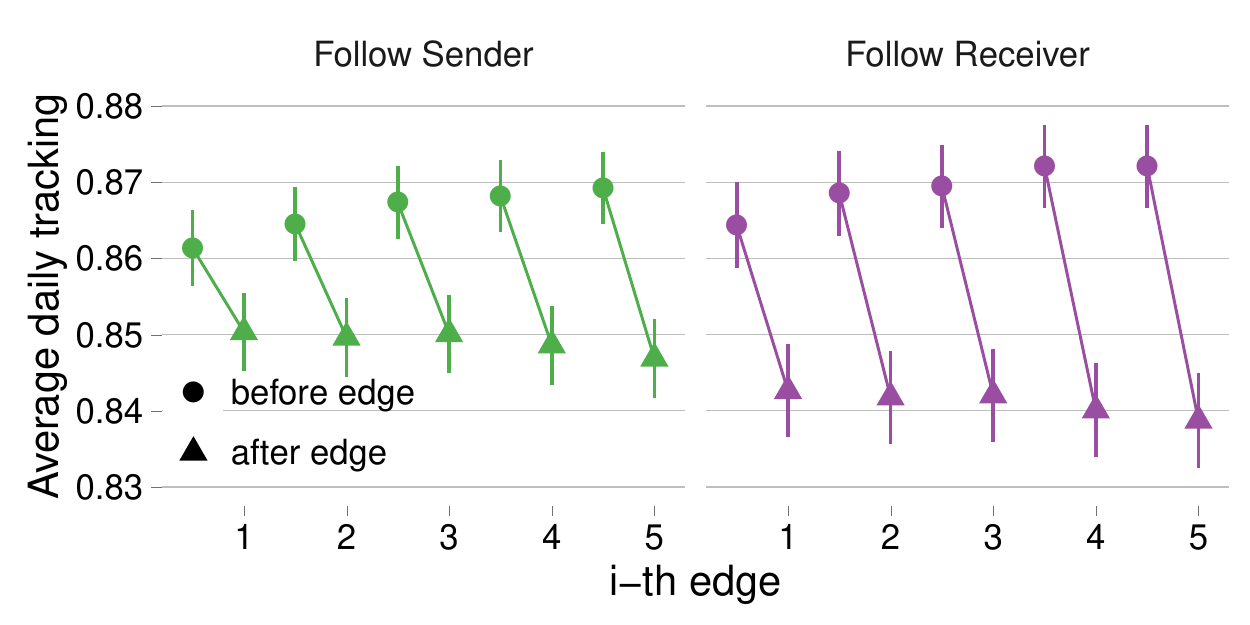}
\caption{
Steps (top) and steps tracking (bottom) before (circles) and after edge creation (triangles) as a function of edge number (x-axis) and edge type and initiator (subplots).
Steps tracking difference refers to the difference in probably of recording steps on a given day of the week after edge creation compared to the week before.
We observe an oscillation pattern of activity levels with decreasing effect sizes (smaller steps differences) at each edge. 
}
\label{fig:edge_effects_oscillating_steps_appendix}
\end{figure*}

The effects of individual edge formations on steps (repeated for comparison) and steps tracking (was not discussed in main paper due to space constraints) are shown in Figure~\ref{fig:edge_effects_steps_wbaseline_appendix} and Figure~\ref{fig:edge_effects_oscillating_steps_appendix}.

The estimated effects of each edge on tracking behavior (\#days with steps tracked each week) are shown in Figure~\ref{fig:edge_effects_steps_wbaseline_appendix} (bottom row), split by edge number (x-axis) and edge type and initiator (subplots) and showing the baselines with dashed lines and rectangles.
We observe very similar results to the activity level differences discussed in Section~\ref{sec:edge_effects}; that is,
we also observe decreasing effects with edge number, larger effects for friend edges than follow edges, and larger effects for the sender of the edge compared to the receiver.
However, in contrast to the previous results where receiving a follower edge was associated with a significant increase in the number of recorded steps, here we observe that sending and receiving follower edges is associated with a significant decrease in the propensity to track steps on a given day.
This suggests that users sending and receiving follower edges are taking more steps in the following week (Figure~\ref{fig:edge_effects_steps_wbaseline_appendix} (top)), but are slightly less likely to track their steps in the first place.

We also note that the steps tracking baseline for friend edges are significantly higher than the baselines for follower edges. 
These dynamics are explained by the timing when friend and follower edges are created.
Friend edges typically happen early in the user's lifetime when tracking (and also activity levels) are increasing on average.
Follower edges happen later in the user's lifetime when, on average, users start tracking steps less regularly.
For example, to receive a follow edge no action is necessary and it is possible that the followed user was about to stop using the app anyway.
For the effect of friend edges on steps tracking we find very similar effects compared to the effect on daily steps including significant increases in the propensity to track steps around the time of the edge creation.

\end{document}